\documentclass[10pt,aps,pra,twocolumn,superscriptaddress,floatfix,showpacs,longbibliography]{revtex4-1}
\usepackage{graphicx}
\usepackage{amsfonts}
\usepackage{amssymb}
\usepackage{amsmath}
\usepackage{txfonts}
\usepackage{lipsum}
\usepackage{color}
\usepackage{wasysym}
\usepackage{hyperref}
\usepackage{bbold}
\usepackage{etoolbox} 

\usepackage{mathtools}
\usepackage{multirow}
\usepackage{hhline}

\DeclareMathOperator{\Tr}{Tr}

\newcommand{\av}[1]{\ensuremath{\left\langle #1 \right\rangle}}
\newcommand{\qv}{\mathbf{q}}
\newcommand{\kv}{\mathbf{k}}

\usepackage{letltxmacro}
\LetLtxMacro{\oldsqrt}{\sqrt}
\renewcommand{\sqrt}[2][\mkern8mu]{\mkern-6mu\mathop{}\oldsqrt[#1]{#2}}

\begin{document}
\title{
Effective Ising model for correlated systems with charge ordering
}
\author{E. A. Stepanov}
\affiliation{Radboud University, Institute for Molecules and Materials, 6525AJ Nijmegen, The Netherlands}
\affiliation{Theoretical Physics and Applied Mathematics Department, Ural Federal University, Mira Street 19, 620002 Ekaterinburg, Russia}

\author{A. Huber}
\affiliation{Institute of Theoretical Physics, University of Hamburg, 20355 Hamburg, Germany}

\author{A. I. Lichtenstein}
\affiliation{Institute of Theoretical Physics, University of Hamburg, 20355 Hamburg, Germany}
\affiliation{Theoretical Physics and Applied Mathematics Department, Ural Federal University, Mira Street 19, 620002 Ekaterinburg, Russia}

\author{M. I. Katsnelson}
\affiliation{Radboud University, Institute for Molecules and Materials, 6525AJ Nijmegen, The Netherlands}
\affiliation{Theoretical Physics and Applied Mathematics Department, Ural Federal University, Mira Street 19, 620002 Ekaterinburg, Russia}

\begin{abstract}
Collective electronic fluctuations in correlated materials give rise to various important phenomena, such as charge ordering, superconductivity, Mott insulating and magnetic phases, and plasmon and magnon modes. Unfortunately, the description of these correlation effects requires significant effort, since they almost entirely rely on strong local and nonlocal electron-electron interactions. Some collective phenomena, such as magnetism, can be sufficiently described by simple Heisenberg-like models that are formulated in terms of bosonic variables. This fact suggests that other many-body excitations can also be described by simple bosonic models in the spirit of Heisenberg theory. Here we derive an effective bosonic action for charge degrees of freedom for the extended Hubbard model and define a physical regime where the obtained action reduces to a classical Hamiltonian of an effective Ising model.
\end{abstract}

\maketitle

\section{Introduction}

Remarkably, the majority of studies of collective charge excitations in modern condensed matter theory is still limited to the random phase approximation (RPA)~\cite{pines1966theory, platzman1973waves, vonsovsky1989quantum}. Although this approach fulfills the charge conservation law and provides a qualitatively good description of plasmonic modes, it is based on a perturbation expansion and, strictly speaking, is applicable only to systems with relatively small Coulomb interaction. A correct description of plasmons in the correlated regime of large electron-electron interactions requires consideration of additional diagrammatic contributions to the electronic self-energy and polarization operator that contain vertex corrections. Unfortunately, the latter implies the use of advanced numerical techniques, which in the case of realistic calculations is extremely time-consuming. Additional diagrammatic corrections often violate the charge conservation law~\cite{PhysRevB.87.125149, PhysRevB.90.235105}, which affects the result for the plasmonic spectrum. Nevertheless, recently a new theory that allows a conserving description of plasmons beyond RPA was proposed in~\cite{PhysRevLett.113.246407}. This approach is based on the Dual Boson (DB) theory~\cite{Rubtsov20121320, PhysRevB.93.045107} and considers the polarization operator in the two-particle ladder form written in terms of local three- and four-point vertex functions. A further extension of this method to the multiorbital case is challenging due to its complicated diagrammatic structure.

Another interesting feature of collective charge excitations in many realistic materials is a tendency of the systems to the charge ordering (CO), which is widely discussed in the literature starting from the discovery of the Verwey transition in magnetite Fe$_3$O$_4$~\cite{VERWEY1941979, doi:10.1063/1.1746466, Mott}. Nowadays, there is a number of other materials, such as the rare-earth compound Yb$_4$As$_3$~\cite{PhysRevB.71.075115, 0295-5075-31-5-6-013,doi:10.1080/0001873021000057114}, transition metal MX$_2$~\cite{PhysRevB.89.235115, ritschel2015orbital, ugeda2016characterization} and rare-earth R$_3$X$_4$~\cite{FURUNO1988117, IRKHIN199047, doi:10.1080/01418638008221893} chalcogenides (${\rm M = V, Nb, Ta}$; ${\rm R = Eu, Sm}$; ${\rm X = S, Se}$), Magn\'eli phase Ti$_4$O$_7$~\cite{doi:10.1080/01418638008221887, doi:10.1080/01418638008221888, EYERT2004151, 0953-8984-18-48-022}, vanadium bronzes Na$_x$V$_2$O$_5$ and Li$_x$V$_2$O$_5$ (see Ref.~\onlinecite{doi:10.1080/0001873021000057114, doi:10.1080/01418638008221890}, and references therein), where the charge ordering has been observed. Since this phenomenon is based on the presence of strong local and nonlocal electron-electron interactions, the theoretical description of this issue also requires the use of very advanced approaches (see e.g. Refs.~\onlinecite{PhysRevB.95.115107, 2017arXiv170705640V}). 

Recent theoretical investigations of charge  correlation effects caused by the strong nonlocal Coulomb interaction indicate that the description of collective charge excitations in the correlated regime can be drastically simplified. Thus, the study of the charge ordering within the dynamical cluster approximation (DCA)~\cite{PhysRevB.95.115149}, Dual Boson~\cite{PhysRevB.90.235135, PhysRevB.94.205110} and GW+EDMFT~\cite{PhysRevB.87.125149, PhysRevB.95.245130} approaches showed similar results for the phase boundary between the normal and CO phases at half filling. The fact that a much simpler GW+EDMFT theory performs in reasonable agreement with the more advanced DB approach and with almost exact DCA method suggests that collective charge fluctuations can be described via a simple theory, at least in a specific physical regime. Unfortunately, the use of the GW+EDMFT theory for description of charge excitations is not fully justified, since this approach suffers from the Fiertz ambiguity when the charge and spin channels are considered simultaneously~\cite{PhysRevB.92.115109, PhysRevB.93.235124}, and from the ``HS-$UV/V$'' decoupling problem~\cite{PhysRevB.66.085120, PhysRevLett.90.086402}. 
In this regard, the simplified (DB$-$GW)~\cite{PhysRevB.90.235135, PhysRevB.94.205110} approximation of the DB theory, which does not consider vertex corrections and is free of the above-mentioned problems, seems more preferable. However, it provides much worse results than the DB~\cite{PhysRevB.94.205110} and GW+EDMFT~\cite{PhysRevB.95.245130} theories. Therefore, the problem of the efficient description of collective charge excitations in correlated materials is still open. 

In the case when accurate quantum mechanical calculations are challenging, the initial quantum problem can be replaced by an appropriate classical one. This thermodynamical approach is widely used, for example, for a description of the ordering in alloys~\cite{PhysRevB.70.125115, PhysRevB.72.104437, PhysRevB.79.054202, PhysRevLett.105.167208, PhysRevB.83.104203, 0034-4885-71-4-046501}. There, the total energy of the ground state is mapped onto an effective Ising Hamiltonian, with parameters determined from {\it ab initio} calculations within the framework of the density functional theory~\cite{PhysRevB.27.5169, 0305-4608-13-11-017, ducastelle1991order}. However, to our knowledge, no attempts to extend this theory to the description of charge fluctuations in the correlated regime and to derive the pair interaction of the Ising model directly from the quantum problem have been reported yet. Additional impulse for investigation of this important problem is given by theoretical studies of magnetism in correlated electronic systems~\cite{LKG84,LKG85,LKAG87,KL2000,PhysRevLett.121.037204}, where an effective classical Heisenberg model for the quantum problem was derived. Since magnetism is also a collective electronic property, one may expect that charge degrees of freedom can be treated in a similar way. 

Motivated by above discussions, we introduce here a new theory that describes charge excitations of the extended Hubbard model in terms of bosonic variables that are related to electronic charge degrees of freedom. The corresponding bosonic action of the model is derived with the use of the advanced ladder DB approach. Consequently, the charge susceptibility has a complicated diagrammatic structure that takes into account frequency dependent vertex corrections. We also observe that the dependence of local vertex functions on fermionic frequencies is directly connected to the value of the double occupancy of lattice sites. Moreover, we find that in a wide range of physical parameters, when the double occupancy is large, this dependence is negligible, and the expression for the charge susceptibility can be drastically simplified. Thus, the theory reduces to an improved version of the GW+EDMFT and DB$-$GW approaches, where the susceptibility takes a simple RPA+EDMFT form. The further application of the derived simple theory to the hole-doped extended Hubbard model shows almost perfect agreement of the obtained result for the phase boundary between the normal and CO phases with much more elaborate and time-consuming ladder DB and DCA~\cite{PhysRevB.97.115117} methods. Finally, it has been shown that in the case of well-developed collective charge fluctuations the initial quantum problem can be mapped onto an effective classical Ising Hamiltonian written in terms of pair interaction between charge densities. This formalism can be efficiently used for the calculation of finite-temperature thermodynamic properties of the system. For instance, we show that the effective Ising model predicts the transition temperature between the normal and charge ordered phases in a good agreement with the DCA result, although our calculations are performed in the unbroken symmetry phase.

\section{Bosonic action for electronic charge} 
Let us start with the following action of the extended Hubbard model written in the Matsubara frequency ($\nu, \omega$) and momentum ($\kv, \qv$) space 
\begin{align}
\label{eq:actionlatt}
{\cal S} = -\sum_{\kv,\nu} c^{*}_{\kv\nu} \left[i\nu+\mu-\varepsilon^{\phantom{*}}_{\kv}\right]
 c^{\phantom{*}}_{\kv\nu} + \frac{1}{2}\sum_{\qv,\omega} \left[U+V_{\qv}\right] \rho^{*}_{\qv\omega}\rho^{\phantom{*}}_{\qv\omega}.
\end{align}
Here $c^{*}_{\kv\nu}$ ($c^{\phantom{*}}_{\kv\nu}$) are Grassmann variables corresponding to the creation (annihilation) of an electron. $\varepsilon_{\kv}$ is the Fourier transform of the hopping amplitude $t_{ij}$, which is considered here in the nearest neighbor approximation on a two-dimensional square lattice. The energy scale is $4t=1$. $U$ and $V_{\qv}$ are local and nonlocal Coulomb interactions, respectively. Charge degrees of freedom are described here introducing the bosonic variable $\rho_{\qv\omega}=n_{\qv\omega}-\av{n_{\qv\omega}}$ that describes variation of the electronic density $n_{\qv\omega} = \sum_{\kv,\nu,\sigma} c^{*}_{\kv\nu\sigma} c^{\phantom{*}}_{\kv+\qv,\nu+\omega,\sigma}$ from the average value. Hereinafter, spin labels $\sigma=\uparrow,\downarrow$ are omitted. 

An effective bosonic action for charge degrees of freedom can be derived following transformations, as presented in a recent work~\cite{PhysRevLett.121.037204}. There, the lattice action~\eqref{eq:actionlatt} is divided into the local impurity problem of the extended dynamical mean-field theory (EDMFT)~\cite{PhysRevB.52.10295, PhysRevLett.77.3391, PhysRevB.61.5184, PhysRevLett.84.3678, PhysRevB.63.115110} and the remaining nonlocal part. In order to decouple the single-electronic and collective charge degrees of freedom, one can perform {\it dual} transformations of the nonlocal part of the lattice action that lead to a new problem written in the {\it dual} space~\cite{PhysRevB.90.235135, PhysRevB.94.205110}. The inverse transformation back to the initial ``lattice'' space after truncation of the interaction of the {\it dual} action at the two-particle level results in the following bosonic action for charge variables (for details see Ref.~\onlinecite{PhysRevLett.121.037204} and Appendix~\ref{App1})
\begin{align}
{\cal S}_{\rm ch}
&=-\frac12\sum_{\qv,\omega} \rho^{*}_{\qv\omega} 
X_{\qv\omega}^{-1} \, \rho^{\phantom{*}}_{\qv\omega}.
\label{eq:Scharge}
\end{align}
Here, the charge susceptibility $X_{\qv\omega}$ in the conserving ladder DB approximation is given by the following relation~\cite{PhysRevLett.121.037204}
\begin{align}
X^{-1}_{\qv\omega} = \left[X^{\rm DMFT}_{\qv\omega}\right]^{-1} + \Lambda_{\omega} - V_{\qv},
\label{eq:Xch}
\end{align}
where $\Lambda_{\omega}$ is the local bosonic hybridization function of the impurity problem. $X^{\rm DMFT}_{\qv\omega} = \sum_{\nu\nu'}\big[X^{\rm DMFT}_{\qv\omega}\big]_{\nu\nu'}$ is the charge susceptibility in the DMFT form~\cite{PhysRevLett.62.324, RevModPhys.68.13} written in terms of lattice Green's functions $G_{\kv\nu}$ and two-particle irreducible (2PI) in the charge channel four-point vertices $\overline{\gamma}^{\,\rm 2PI}_{\nu\nu'\omega}$ of the local impurity problem (see Appendix~\ref{App1})
\begin{align}
\left[X^{\rm DMFT}_{\qv\omega}\right]^{-1}_{\nu\nu'} = 
\left[X^{0}_{\qv\omega}\right]^{-1}_{\nu\nu'} + \overline{\gamma}^{\,\rm 2PI}_{\nu\nu'\omega}.
\end{align}
Here, $\big[X^{0}_{\qv\omega}\big]_{\nu\nu'}=\sum_{\kv}G_{\kv+\qv,\nu+\omega}G_{\kv\nu}\, \delta_{\nu\nu'}$ is a generalized bare lattice susceptibility, and the inversion should be understood as a matrix operation in the fermionic frequency $\nu,\nu'$ space. 
Note that in the ladder DB approximation the lattice Green's function is dressed only in the local impurity self-energy and therefore coincides with the usual EDMFT expression~\cite{PhysRevB.52.10295, PhysRevLett.77.3391, PhysRevB.61.5184, PhysRevLett.84.3678, PhysRevB.63.115110}. Thus, the relation for the lattice susceptibility can be written as $X^{\phantom{1}}_{\qv\omega} = \sum_{\nu\nu'}\big[X^{\phantom{1}}_{\qv\omega}\big]_{\nu\nu'}$, where 
\begin{align}
\left[X^{\phantom{1}}_{\qv\omega}\right]_{\nu\nu'}^{-1} = \left[X^{0}_{\qv\omega}\right]_{\nu\nu'}^{-1} - U^{\rm eff}_{\nu\nu'\omega} - V_{\qv},
\label{eq:Xch2}
\end{align}
and we introduced an effective bare local Coulomb interaction 
\begin{align}
U^{\rm eff}_{\nu\nu'\omega} = -\Lambda_{\omega} - \overline{\gamma}^{\,\rm 2PI}_{\nu\nu'\omega}.
\label{eq:Uvvw}
\end{align}
Note that the 2PI vertex function $\overline{\gamma}^{\,\rm 2PI}_{\nu\nu'\omega}$ is defined here in the particle-hole channel.

A recent study of magnetism of correlated electrons~\cite{PhysRevLett.121.037204} shows that if the system exhibits well-developed bosonic fluctuations, the corresponding local vertex functions mostly depend on bosonic frequency $\omega$, while their dependence on fermionic frequencies $\nu,\nu'$ is negligible. Therefore, one can expect that in a physical regime where charge fluctuations are dominant the local 2PI vertex function in the charge channel can be approximated as $\overline{\gamma}^{\rm 2PI}_{\nu\nu'\omega}\simeq\overline{\gamma}^{\rm 2PI}_{\omega}$, and the charge susceptibility~\eqref{eq:Xch} takes the following simple form 
\begin{align}
X^{-1}_{\qv\omega} = X^{0~-1}_{\qv\omega} - \left(U^{\rm eff}_{\omega} + V_{\bf q}\right).
\label{eq:RPA}
\end{align}
Here, $X^{0}_{\qv\omega}=\sum_{\nu\nu'}\big[X^{0}_{\qv\omega}\big]_{\nu\nu'} = \sum_{\kv\nu}G_{\kv+\qv,\nu+\omega}G_{\kv\nu}$ is the bare lattice susceptibility, and the effective bare local Coulomb interaction~\eqref{eq:Uvvw} transforms to $U^{\rm eff}_{\omega} = -\Lambda_{\omega} - \overline{\gamma}^{\,\rm 2PI}_{\omega}$. As it is also shown in Ref.~\onlinecite{PhysRevLett.121.037204} and Appendix~\ref{App3}, in the considered case of well-developed collective fluctuations the 2PI vertex function can be approximated as 
\begin{align}
\gamma^{\rm 2PI}_{\omega}\simeq\chi^{-1}_{\omega}-\chi^{0~-1}_{\omega}\simeq-U-\Lambda_{\omega},
\end{align}
where $\chi_{\omega}$ and $\chi^{0}_{\omega}$ are the full and bare local susceptibilities of the impurity problem, respectively. As a consequence, the effective bare local Coulomb interaction reduces to the actual value of the local Coulomb interaction $U^{\rm eff}_{\omega}\simeq{}U$. Therefore, the expression in Eq.~\ref{eq:RPA} is nothing more than the RPA susceptibility constructed on top of the EDMFT result for Green's functions. This simplified approximation is referred in the text to as the RPA+EDMFT approach. 

It is worth noting that in the regime of strong charge fluctuations the local self-energy takes the same form as in GW approach~\cite{PhysRev.139.A796, 0034-4885-61-3-002, 0953-8984-11-42-201} (see Ref.~\onlinecite{PhysRevLett.121.037204} and Appendix~\ref{App3}). Hence, the simplified theory can be reduced to the GW method in the case when the nonlocal contribution to the self-energy is also considered. Thus, we show that it is indeed possible to describe strong charge excitations by a simple bosonic action~\eqref{eq:Scharge} in terms of charge susceptibility~\eqref{eq:RPA} that does not contain vertex corrections. 

\begin{figure}[t!]
\includegraphics[width=0.8\linewidth]{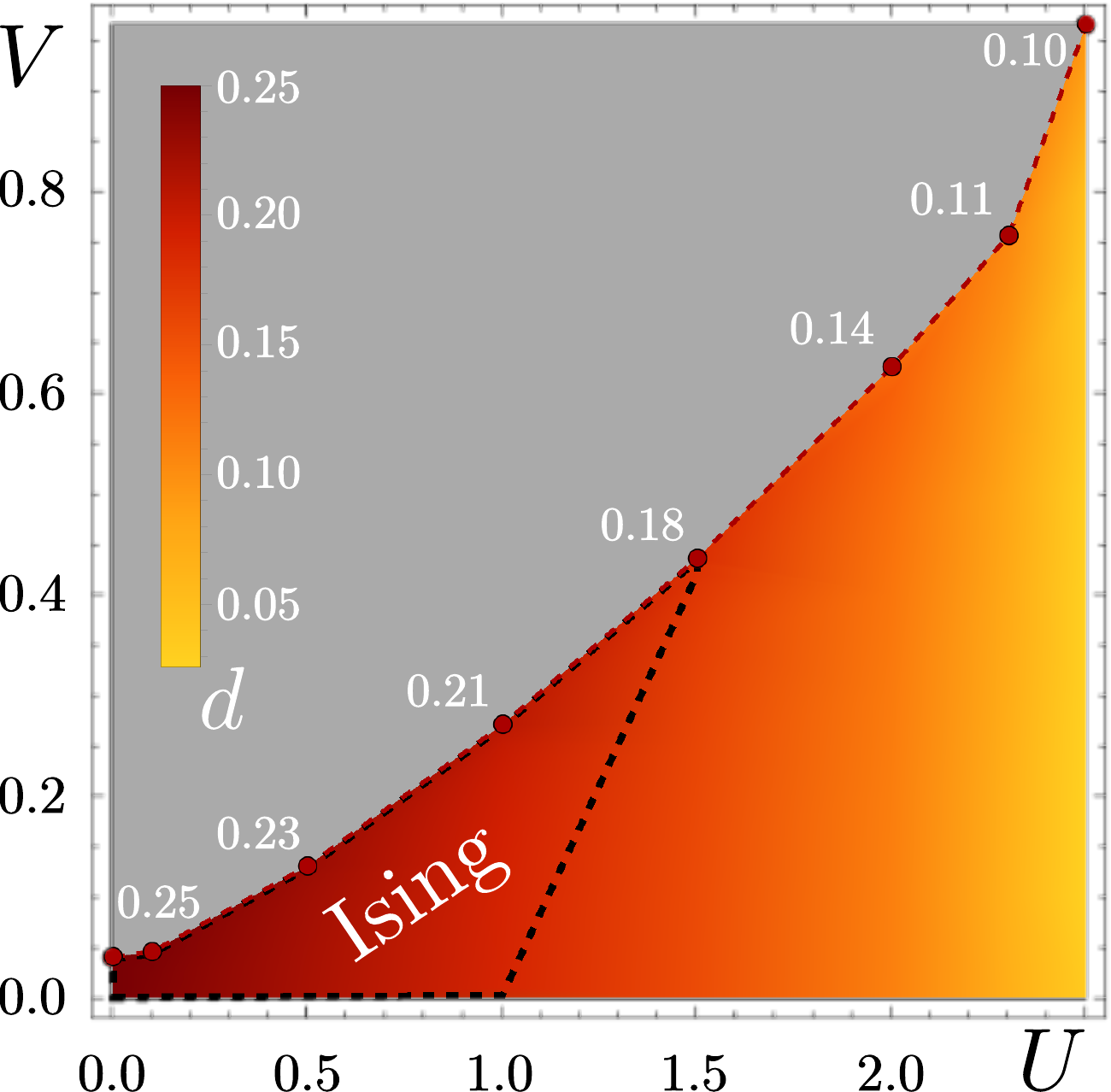}
\caption{(Color online) Double occupancy of the extended Hubbard model shown on the $U$-$V$ phase diagram. Calculations are performed in the normal phase where the value of the double occupancy $d$ is depicted by color. The gray part corresponds to the charge ordered phase. Values of the double occupancy at the phase boundary are explicitly mentioned. The area depicted by the black dashed line corresponds to the case of large value of the double occupancy $d\gtrsim70\%\,d_{\rm max}$ and shows the regime where charge excitations can be described by an effective Ising model. Values of Coulomb interactions $U$ and $V$ are given in units of half of the bandwidth ($W/2=4t=1$). Therefore, the effective Ising model can be used for a broad range of values of the Coulomb interaction, which may even exceed half of the bandwidth. The inverse temperature is $\beta=50$. \label{fig:DO}}
\end{figure}

\section{Regime of strong charge fluctuations}

Now, let us define the physical regime where the presented above technique is applicable. In Ref.~\onlinecite{PhysRevLett.121.037204} collective excitations have been studied in the ordered (antiferromagnetic) phase, where the proximity of the local magnetic moment $m$ to its maximum value served as a signature of well-developed spin fluctuations. Here, we are interested in a similar description of a more complicated case when collective charge excitations are present in the system already in the normal phase. Since in the latter case all lattice sites are described by the same local impurity problem, the corresponding signature of strong bosonic fluctuations can no longer be found among local single-particle observables that are identical for every lattice site. It is worth mentioning that, contrary to the magnetic phase where the ordering of single-particle quantites (local magnetizations) is realized, the CO phase on a lattice corresponds to the ordering of dublons (see i.e. Refs.~\onlinecite{0022-3719-12-11-015} and~\onlinecite{PhysRevB.56.12939}) that are two-particle observables. Thus, the double-occupancy of the lattice site, which is defined as $d=\av{n_{\uparrow}n_{\downarrow}}$ with the maximum value $d_{\rm max}=0.25$ in the normal phase, can be proposed as a fingerprint of the existence of strong charge fluctuations in the system. 

The corresponding result for the double occupancy of the two-dimensional extended Hubbard model~\eqref{eq:actionlatt} on the square lattice is shown on the $U$-$V$ phase diagram in Fig.~\ref{fig:DO} and obtained using the DB approach~\cite{PhysRevB.93.155162} without the approximation of the four-point vertex function inrodused above. The phase boundary (red dashed line) between the normal (colored) and CO (gray) phases is determined from the zeros of the inverse charge susceptibility $X^{-1}_{\qv\omega}$~\eqref{eq:Xch} at $\qv=(\pi, \pi)$ and $\omega=0$ point similarly to Refs.~\onlinecite{PhysRevB.90.235135, PhysRevB.94.205110}. As already mentioned in the Introduction, this result for the phase boundary is in a very good agreement with the DCA calculations performed in Ref.~\onlinecite{PhysRevB.95.115149}. As expected, large charge fluctuations in the normal phase emerge in the region close to the phase transition to the ordered state. However, one can see that the strength of these fluctuations is not uniformly distributed along the phase boundary, since the value of $d$ decreases with the increase of the local Coulomb interaction. 

\begin{figure}[t!]
\includegraphics[width=0.8\linewidth]{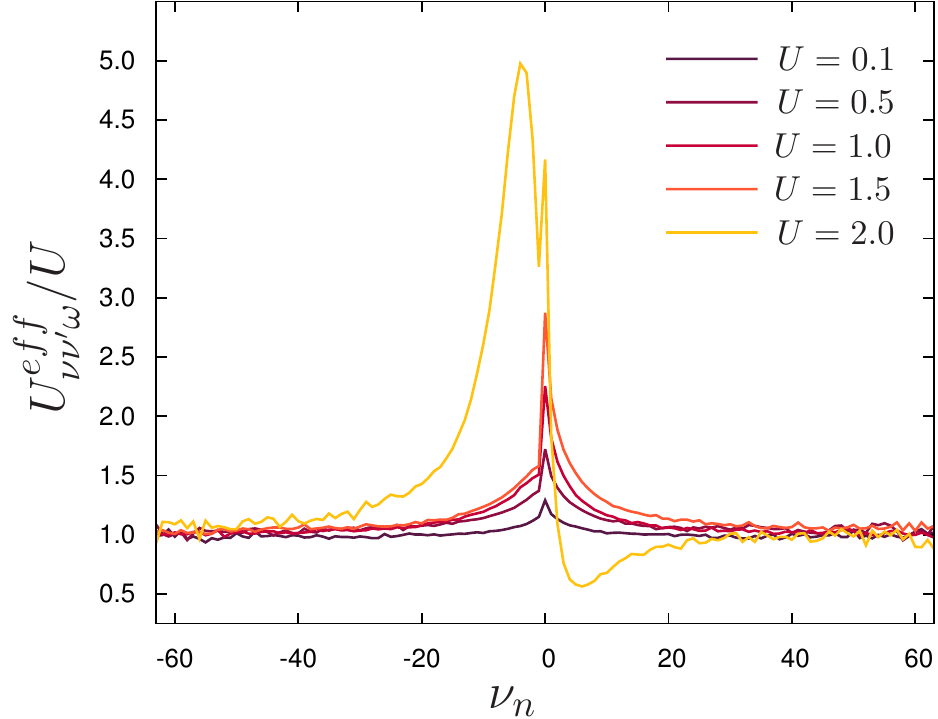}
\caption{(Color online) Frequency dependence of the effective local Coulomb interaction $U^{\rm eff}_{\nu\nu'\omega}$ obtained for different values of $U$ at the phase boundary between the normal and CO phases at the $\nu'=0$ and $\omega=0$ point for $\beta=50$. As the double occupancy $d$ is decreased, the dependence of the effective interaction $U^{\rm eff}_{\nu\nu'\omega}$ on fermionic frequency becomes larger.\label{fig:vertex}}
\end{figure}

In order to clarify the connection between the value of the double occupancy and the strength of charge fluctuations, one can study an effective bare local Coulomb interaction $U^{\rm eff}_{\nu\nu'\omega}$ defined in Eq.~\ref{eq:Uvvw}. Fig.~\ref{fig:vertex} shows the ratio $U^{\rm eff}_{\nu\nu'\omega}/U$ between the effective and actual local Coulomb interactions as the function of fermionic frequency $\nu$ at the $\nu'=\omega=0$ point. This result is obtained close to the phase boundary between the normal and CO phases shown in Fig.~\ref{fig:DO} for different values of the local Coulomb interaction $U$ and, as a consequence, of the double occupancy $d$. The exact values of $U$, $V$, and $d$ for these calculations are specified in Table~\ref{table}. Here, one can immediately see that the effective Coulomb interaction at small values of $U$ (large values of $d$) is almost frequency independent. Decreasing the double occupancy, the frequency dependence of $U^{\rm eff}$ becomes crucial, and one can no longer approximate the local 2PI vertex function by neglecting its dependence on fermionic frequencies. Remarkably, the effective Coulomb interaction tends to the actual value of the local Coulomb interaction at large frequencies for every value of $U$, which is in perfect agreement with the theory presented above. A similar asymptotic behavior was reported for the 2PI vertex function of the DMFT impurity problem ($\Lambda_{\omega}=0$) in Ref.~\onlinecite{PhysRevB.86.125114}. Thus, one can conclude that the presence of the bosonic hybridization function $\Lambda_{\omega}$ in the local impurity problem changes local vertex functions. The presence of $\Lambda_{\omega}$ in Eq.~\ref{eq:Xch} restores the correct frequency behavior of the lattice susceptibility by cancelling the bosonic hybridization from the vertex function in the effective local interaction. Therefore, the inclusion of the $\Lambda_{\omega}$ in the theory has to be done consistently both in the local impurity problem and the lattice susceptibility~\eqref{eq:Xch}, otherwise it may lead to incorrect frequency behavior of bosonic quantities. Results for $U^{\rm eff}_{\nu\nu'\omega}/U$ for other values of $\nu'$ and $\omega$ can be found in Appendix~\ref{App1} and show a similar connection of the double occupancy to the frequency dependence of the effective Coulomb interaction.

\begin{figure}[t!]
\includegraphics[width=0.8\linewidth]{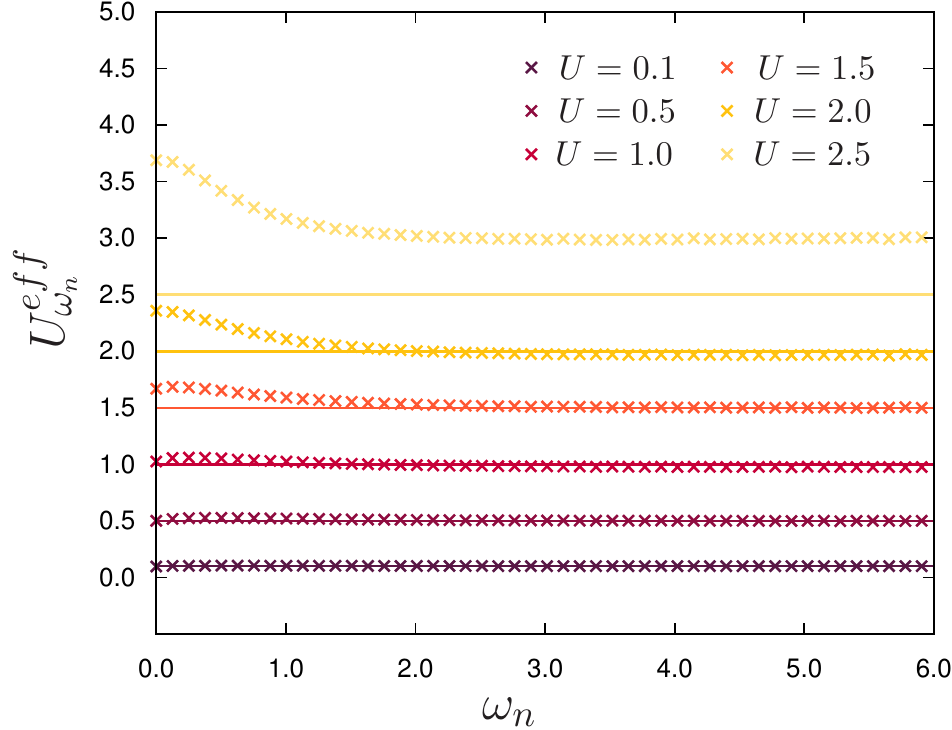}
\caption{(Color online) Frequency dependence of the effective local Coulomb interaction $U^{\rm eff}_{\omega}$ obtained for $\beta=50$ close to the phase boundary between the normal and charge ordered phases for different values of the actual Coulomb interaction $U$. When the double occupancy is decreased, the difference between the effective and actual local Coulomb interactions becomes more notable. \label{fig:Ueff}}
\end{figure}

Let us now investigate the dependence of the effective local Coulomb interaction on the bosonic frequency $\omega$. As shown in Fig.~\ref{fig:vertex}, the use of the fermionic frequency independent approximation $\overline{\gamma}^{\rm 2PI}_{\nu\nu'\omega}\simeq\overline{\gamma}^{\rm 2PI}_{\omega}$ for the 2PI vertex in the large double occupancy regime is now justified. Then, the effective Coulomb interaction $U^{\rm eff}_{\omega}$ can be extracted from the simplified expression for the charge susceptibility~\eqref{eq:RPA}, where the left-hand side is substituted from Eq.~\ref{eq:Xch}. Since the leading contribution to the lattice susceptibility in this regime is given by the $\qv=(\pi, \pi)$ momentum, the corresponding effective interaction shown in Fig.~\ref{fig:Ueff} reads 
\begin{align}
U^{\rm eff}_{\omega} = X^{0~-1}_{(\pi,\pi),\omega} - \left[X^{\rm DMFT}_{(\pi,\pi),\omega}\right]^{-1} - \Lambda_{\omega}.
\end{align}
Here, the result is obtained in the normal phase close to the CO for the same values of Coulomb interactions as in Fig.~\ref{fig:vertex}. 
It is worth mentioning that the above definition of the effective local Coulomb interaction is similar to the one of the two-particle self-consistent theory proposed by Vilk and Tremblay~\cite{refId0}. However, we use a more advanced ladder DB expression~\eqref{eq:Xch} for the lattice susceptibility, contrary to the RPA form with bare Green's functions considered in their work.

Remarkably, when the double occupancy is close to its maximum value, the effective Coulomb interaction $U^{\rm eff}$ does not depend on bosonic frequency either, and again coincides with the actual Coulomb interaction. In the smaller $d$ regime the bosonic frequency dependence appears and cannot be avoided for consideration anymore. Therefore, the large value of the double occupancy is indeed an indicator of a well-developed charge fluctuations. Taking into account results shown in Figs.~\ref{fig:vertex} and~\ref{fig:Ueff}, the value of the double occupancy for which the effective local interaction is frequency independent and coincides with the bare local Coulomb interaction $U$ can be estimated as $d\gtrsim70\%\,d_{\rm max}$. As schematically shown in Fig.~\ref{fig:DO} by the black dashed line, the corresponding region where the use of a simple RPA+EDMFT approach is justified can be distinguished for the relatively broad range of Coulomb interactions. Surprisingly, the latter may even exceed half of the bandwidth. 

\section{Extended Hubbard model upon doping}

\begin{figure}[t!]
\includegraphics[width=0.95\linewidth]{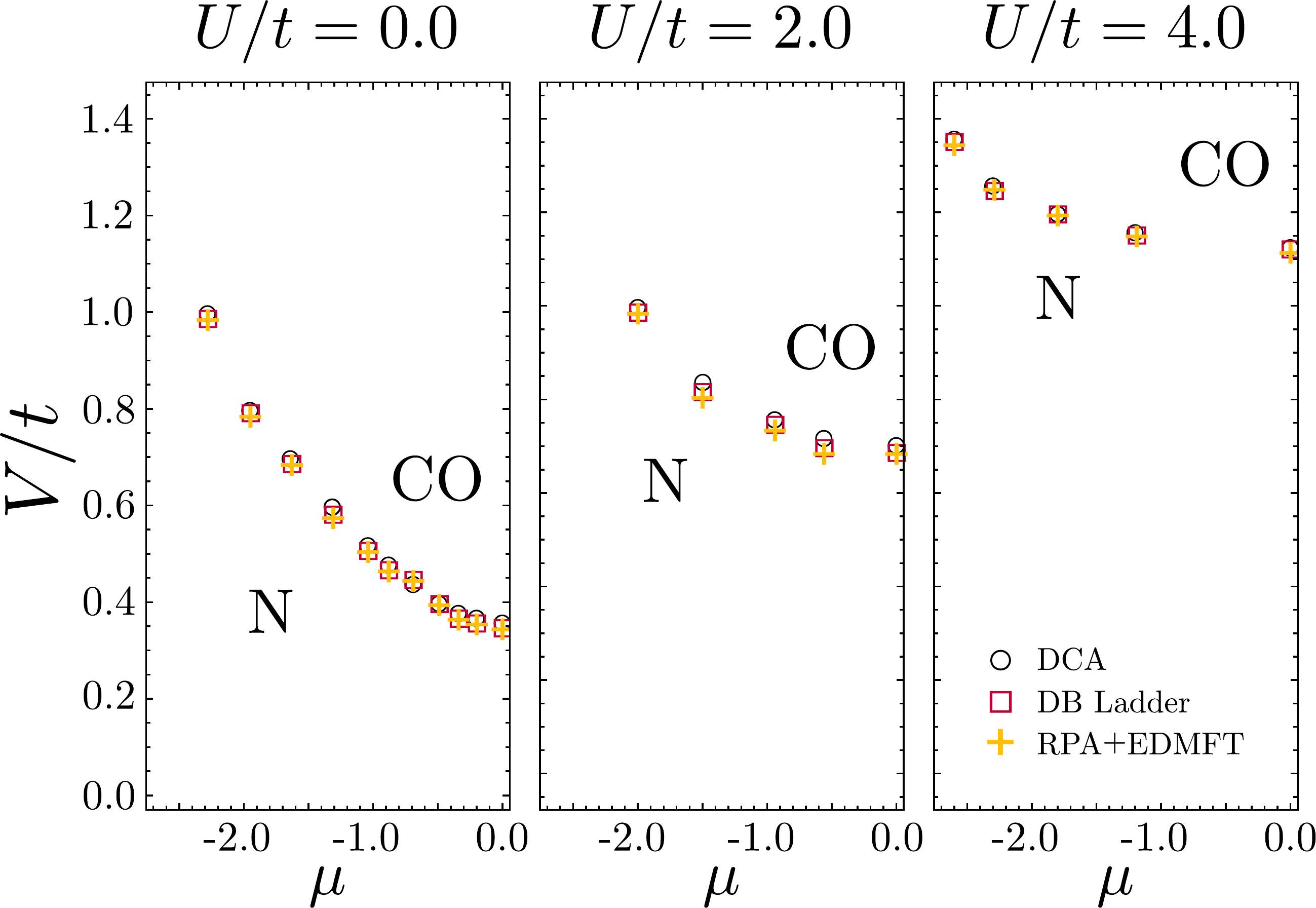}
\caption{(Color online) Phase boundary between the normal (N) and charge-ordered (CO) phases of the hole-doped extended Hubbard model in the space of nearest-neighbor interaction $V$ and chemical potential $\mu$. Calculations are performed for the Ladder DB and RPA+EDMFT approaches in the regime of large double occupancy where the use of the simplified approximation is justified. The DCA data is kindly provided by authors of the Ref.~\onlinecite{PhysRevB.97.115117}. The local interaction is $U=0$ (left panel), $U=0.5$ (middle panel), and $U=1$ (right panel). The hopping amplitude is $t=0.25$. $\mu=0$ corresponds to the half filling. All data are obtained for $\beta=12.5$ ($T=0.08$). \label{fig:dop}}
\end{figure}

Calculation of phase boundaries became a standard test for the performance of the introduced theory~\cite{PhysRevB.87.125149, PhysRevB.90.235135, PhysRevB.94.205110, PhysRevB.95.115149, PhysRevB.97.115117}. In order to demonstrate the power of the derived above RPA+EDMFT approach in description of strong charge fluctuations, let us investigate the phase boundary between the normal and CO phases of the extended Hubbard model beyond the half filling. Recently, this issue has been addressed with the use of the dynamical cluster approximation~\cite{PhysRevB.97.115117} in the hole-doped case. DCA is a very advanced approach, which is based on a cluster dynamical mean field theory. Thus, the result obtained in Ref.~\onlinecite{PhysRevB.97.115117} for the phase boundary can be considered as a benchmark. In the previous section we have distinguished the physical regime of applicability of the RPA+EDMFT approach at the half filling. This region is depicted by the black dashed line in Fig.~\ref{fig:DO}. In order to study the performance of the RPA+EDMFT approach against more advanced ladder DB and DCA theories, we obtain the phase boundary in the same region of physical parameters upon the hole doping. The corresponding result is shown in Fig.~\ref{fig:dop} in the space of nearest-neighbor interaction $V$ and chemical potential $\mu$ for $U=0$ (left panel), $U=0.5$ (middle panel), and $U=1$ (right panel). The value of the chemical potential is counted from the half filling ($\mu=0$). The temperature $T = 0.08$ ($\beta=12.5$) for numerical calculations is taken the same as in Ref.~\onlinecite{PhysRevB.97.115117}. The RPA+EDMFT result (yellow pluses) for the phase boundary is obtained using the expression~\eqref{eq:RPA} for the charge susceptibility, where the effective local interaction $U^{\rm eff}_{\omega}$ is replaced by the actual value of the Coulomb interaction $U$ according to above discussions. The ladder DB result (red squares) is obtained using the Eq.~\ref{eq:Xch} as a single shot calculation on top of the converged EDMFT solution. The DCA data (black circles) is kindly provided by authors of Ref.~\onlinecite{PhysRevB.97.115117}. 

The Fig.~\ref{fig:dop} shows that results for the phase boundary between the normal (N) and charge ordered (CO) phases almost perfectly coincide for all three theories for different values of local Coulomb interaction. Remarkably, the result of RPA+EDMFT is in a good agreement with more elaborate methods even at large values of doping. Thus, results for the phase boundary have been compared up to 17\%, 18\%, and 20\% of hole-doping for $U=0$, $U=0.5$, and $U=1$, respectively. This fact is even more surprising, since the RPA+EDMFT  operates only with Green's functions of the single-site EDMFT solution of the problem, while the Dual Boson approach requires a calculation of local vertex functions in order to perform the diagrammatic extension of the EDMFT. The converged EDMFT solution for a one point in parameter space can be obtained, for example, within 20 iterations (20 minutes each) on a single node of the North-German Supercomputing Alliance (HLRN) cluster. At the same time, the simplest single shot ladder DB calculation requires additional iteration that is to be performed on the same cluster already on 4 nodes (each node contains 24 cores), which takes at least 60 more minutes for the one point.  

Multiorbital version the Dual Boson theory is much more time-consuming, since it requires numerical calculation of vertex functions in the enlarged parameter space, and is not yet implemented. The extension of the DCA method to the multiorbital case is even more complicated. An addition difficulty here corresponds to the fact that DCA calculations cannot be performed at reasonably low temperatures beyond the half filling due to the sign problem. For this reason, the comparison between three theories has been performed at $\beta=12.5$, while previous DB calculations were done for much lower temperature $\beta=50$. Therefore, the RPA+EDMFT appears to be a very appealing approach for the description of strong collective excitations in the multiorbital case, since it does not require complicated numerical efforts other than the EDMFT solution of the problem.

\section{Effective Ising model} 

In general, the existence of separate dynamics and a corresponding classical Hamiltonian for charge degrees of freedom is questionable. The possibility to introduce a classical problem for certain collective excitations is usually related to the existence of an adiabatic parameter that distinguishes these excitations from others that belong to different energy and time scales. Thus, in the case of spin fluctuations the adiabatic approximation is intuitive and implies that collective (spin) degrees of freedom are slower and have lower energy than single-particle (electronic) excitations~\cite{PhysRevLett.75.729}. Unfortunately, the corresponding adiabatic approximation for charge degrees of freedom does not exist. Therefore, it is very challenging to find a specific physical regime where the classical problem for charge degrees of freedom can still be introduced. As was recently obtained for spin fluctuations~\cite{PhysRevLett.121.037204}, the possibility of different energy and time scales separation lies in a nontrivial frequency behavior of local vertex functions. If the dependence of the local vertex on fermionic (single-particle) frequencies is negligibly small compared to the bosonic (collective) frequency dependence, the separation of the corresponding bosonic excitation is justified. 

Thus, in the regime of the large value of the double occupancy ($d>70\%\,d_{\rm max}$), which is shown in Fig.~\ref{fig:DO} by the dashed black line, the quantum action~\eqref{eq:Scharge} can be mapped onto an effective classical Hamiltonian, similarly to the case of collective spin fluctuations with the well-defined local moment~\cite{PhysRevLett.121.037204}. Note that in the case of charge degrees of freedom, the classical problem is given by the effective Ising Hamiltonian 
\begin{align}
H_{\rm ch} = -\sum_{\qv} J_{\qv}\,\boldsymbol{\sigma}_{\qv}\,\boldsymbol{\sigma}_{-\qv}
\end{align}
written in terms of classical variables $\boldsymbol{\sigma}=\pm1$. An effective pair interaction $J_{\qv}$ between electronic densities can be defined from the nonlocal part of the inverse charge susceptibility at the zero bosonic frequency~\cite{KL2000,PhysRevLett.121.037204}. Additionally, quantum variables $\rho^{(*)}_{\qv\omega}$ that describe a deviation of the local electronic density from the average (half-filled) value have to be replaced in Eq.~\ref{eq:Scharge} as follows $\rho^{*}_{\qv\omega}\rho^{\phantom{*}}_{\qv\omega}\to2d\,\boldsymbol{\sigma}_{\qv}\,\boldsymbol{\sigma}_{-\qv}$. In order to distinguish local and nonlocal contributions to the inverse susceptibility~\eqref{eq:Xch}, one can again use an approximated version of the local 2PI vertex function in the charge channel. Since the latter does not depend on fermionic frequencies in the regime of well-developed charge fluctuations, the full four-point vertex $\overline{\gamma}_{\nu\nu'\omega}$ of the impurity problem can also be approximated by the leading bosonic contribution. According to the Ref.~\onlinecite{PhysRevLett.121.037204} the latter corresponds to the full local charge susceptibility $\chi_{\omega}$ that connects two three-point vertex functions $\gamma_{\nu\omega}$ (for details see Appendix~\ref{App1})
\begin{table}[t!]
\vspace{-0.2cm}
\caption{\label{table} Double occupancy $d$, correction $U'$ to the effective local Coulomb interaction $U^{\rm eff}$, and static dielectric function $\varepsilon$ obtained close to the phase boundary between the normal and CO phases for the given values of the local $U$ and nonlocal $V$ Coulomb interactions.}
\begin{center}
{\renewcommand{\arraystretch}{1.3}
\tabcolsep=5pt
\begin{tabular}{|c||c||c||c||c||c||c|}
\hline
$U$ & 0.1 & 0.5 & 1.0 & 1.5 & 2.0 & 2.5 \\
\hline
$V$ & 0.045 & 0.130 & 0.265 & 0.420 & 0.630 & 0.965 \\
\hline
$d$ & $0.25$ & $0.23$ & $0.21$ & $0.18$ & $0.14$ & $0.10$ \\
\hline
$U'$ & $-0.48$ & $-0.68$ & $-1.11$ & $-1.81$ & $-2.85$ & $-5.24$ \\
\hline
$\varepsilon$ & $1.26$ & $3.78$ & $10.09$ & $6.00$ & $3.35$ & $1.91$ \\
\hline
\end{tabular}
}
\end{center}
\end{table}
\begin{align}
\label{eq:vert}
\overline{\gamma}_{\nu\nu'\omega}\simeq-\gamma_{\nu\omega}\,\chi_{\omega}\,\gamma_{\nu'+\omega,-\omega} = \vcenter{\hbox{\includegraphics[width=0.16\linewidth]{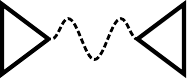}}}\,.
\end{align}
Then, the relation~\eqref{eq:Xch} for the charge susceptibility reduces to 
\begin{align}
X^{-1}_{\qv\omega} = \chi^{-1}_{\omega} + \Lambda_{\omega} - V_{\qv} - \tilde{\Pi}^{(2)}_{\qv\omega},
\end{align}
where the second order polarization operator reads
\begin{align}
\tilde{\Pi}^{(2)}_{\qv\omega} = \sum_{\kv\nu}\gamma_{\nu+\omega,-\omega}\tilde{G}_{\kv+\qv,\nu+\omega} \tilde{G}_{\kv\nu}\gamma_{\nu,\omega},
\end{align}
and $\tilde{G}_{\kv\nu}$ is a nonlocal part of the lattice Green's function. Then, the effective pair interaction takes the following form
\begin{align}
\label{eq:pair}
J_{\qv}/d &= -V_{\qv} - \sum_{\kv,\nu}\gamma_{\nu,0}\tilde{G}_{\kv+\qv,\nu} \tilde{G}_{\kv\nu}\gamma_{\nu,0} \\
&= -V_{\qv} - \vcenter{\hbox{\includegraphics[width=0.18\linewidth]{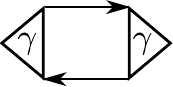}}}. \notag
\end{align}
Using an exact relation between the 2PI four-point and full three-point vertices, the latter can also be approximated as
\begin{align}
\label{eq:Ucorr}
\gamma_{\nu\omega} \simeq \chi^{-1}_{\omega} + \Lambda_{\omega} + U^{\rm eff}_{\nu\nu'\omega} \simeq \chi^{0~-1}_{\omega},
\end{align}
as shown in Appendix~\ref{App3}. Therefore, the result for the pair interaction~\eqref{eq:pair} between electronic densities at first glance reduces to a similar expression for the exchange interaction derived for the magnetic system in Ref.~\onlinecite{PhysRevLett.121.037204}. However, the ``correction'' 
\begin{align}
U'=\chi^{-1}_{\omega=0} + \Lambda_{\omega=0}
\end{align}
to the effective bare Coulomb interaction $U^{\rm eff}$ in expression~\eqref{eq:Ucorr} is larger than the local Coulomb interaction $U$ as shown in Table~\ref{table}. This is not surprising, because the relatively large value of the inversed local charge susceptibility, which is defined as $\chi_{\omega} = -\av{n^{*}_{\omega}\,n^{\phantom{*}}_{\omega}}$, when two electrons occupy the same lattice site is expected. Therefore, the term $U'$ cannot be neglected, contrary to the case of spin fluctuations at half-filling when the inversed local magnetic susceptibility $\chi^{-1}_{\omega=0}$ is negligibly small~\cite{PhysRevLett.121.037204}. Since the effective bare Coulomb interaction $U^{\rm eff}$ in the regime of large double occupancy coincides with the actual value of $U$, one can obtain a static approximation for the three-point vertex (see Appendix~\ref{App3})
\begin{align}
\gamma_{\nu,0}\simeq \chi^{0~-1}_{\omega=0} \simeq -\frac{U}{\varepsilon-1} = -\tilde{U},
\end{align}
where $\varepsilon = U/{\cal W}_{0}$ is a static dielectric function defined via the renormalized local interaction ${\cal W}_{\omega}$. Therefore, the final expression for the pair interaction of the effective classical Ising model reads
\begin{align}
J_{\qv}/d = -V_{\qv} - \sum_{\kv,\nu}\tilde{U}\,\tilde{G}_{\kv+\qv,\nu}\, \tilde{G}_{\kv\nu}\,\tilde{U}.
\label{eq:J}
\end{align}

The effective Ising model can be used for modeling finite-temperature thermodynamic properties
of the system, such as the electronic density, charge susceptibility, ground-state energy, and configurational structure of material~\cite{PhysRevB.70.125115, PhysRevB.72.104437, PhysRevB.79.054202, PhysRevLett.105.167208, PhysRevB.83.104203, 0034-4885-71-4-046501}. All these observables make sense in the broken symmetry (CO) phase. These calculations are beyond the scope of the current paper. However, the Ising model also provides an analytical result for the transition temperature  $T_{\rm c}$ between the normal and CO phases
\begin{align}
T_{\rm c} = 2J / \ln\left(\sqrt{2}+1\right),
\end{align}
where $J=J_{\qv=(\pi,\pi)}/4$ approximates the nearest-neighbor pair interaction. The result for the transition temperature can be compared to the one of the Ref.~\cite{PhysRevB.97.115117} (Fig.~3). To this aim, we obtain the effective exchange interaction at $U=0.5$ for the same values of the nonlocal Coulomb interaction $V=0.19$ and $V=0.275$ used in Ref.~\cite{PhysRevB.97.115117}. Transition temperatures obtained within the DCA in these two cases are $T_c=0.103$ ($\beta\simeq9.69$) and $T_c=0.204$ ($\beta\simeq4.90$), respectively. Since we perform our calculations in the normal phase, the effective exchange interaction was obtained above the critical temperature at $\beta=8$ and $\beta=3$, respectively. These temperatures still allow to get reasonably large values of the double occupancy in order to justify the use of the effective Ising model. Estimated critical temperatures in our calculations are $T_{c}=0.114$ ($J=0.050$, $d=0.230$) and $T_{c}=0.190$ ($J=0.084$, $d=0.247$) respectively, which is in a good agreement with corresponding DCA results. Note that our calculations were performed in the unbroken symmetry phase. We believe that the agreement for the critical temperature is much better for calculations performed in the charge-ordered phase, where the value of the double occupancy is larger and collective fluctuations are much stronger.

\section{Conclusion}

In this work the bosonic action~\eqref{eq:Scharge} for charge degrees of freedom of the extended Hubbard model~\eqref{eq:actionlatt} has been derived. It was found that local four-point vertex function of the impurity model is independent on fermionic frequencies in the regime of well-developed charge fluctuations. Remarkably, the latter can be efficiently determined looking at the deviation of the double occupancy from its maximum value. Thus, strong charge fluctuations are revealed in the case of large double occupancy ($d\gtrsim70\%\,d_{\rm max}$), which corresponds to a broad range of values of Coulomb interaction. As a consequence, it was found that in this regime the dynamics of charge fluctuations can be described via a simplified RPA+EDMFT charge susceptibility~\eqref{eq:RPA} constructed from the EDMFT Green's functions. Moreover the effective local Coulomb interaction in this case coincides with the actual value of the bare Coulomb interaction. Remarkably, the RPA+EDMFT theory performs in a good agreement with more advanced methods even beyond the half filling. Thus, this simple approach correctly predicts the phase boundary between the normal and CO phases up to 20\% of hole doping in a broad range of values of Coulomb interaction. The latter can even reaching the half of the bandwidth ($U/t=4$). Further, it was shown that in the regime of well-developed charge fluctuations, the initial quantum problem can be mapped onto an effective classical Ising model written in terms of a pair interaction between local electronic densities. This is a nontrivial result,
since collective charge excitations cannot be separated from single electronic ones in the same was as it is usually done for spins, because the corresponding adiabatic approximation for charge degrees of freedom does not exist. Importantly, the expression for the pair interaction contains only single-particle quantities, which can be efficiently used in realistic multiband calculations. The predicted critical temperature of the effective Ising model is in a good agreement with the one of the DCA result, which allows to believe that this simple model can be efficiently used for calculation of finite-temperature thermodynamic properties of the system in the ordered phase. We further speculate that similar approximations are valid for realistic multiband systems that reveal strong charge fluctuations. 
\\

\begin{acknowledgments}
We are very thankful to authors of the Ref.~\onlinecite{PhysRevB.97.115117}:  Hanna Terletska, Tianran Chen, Joseph Paki, and Emanuel Gull, for sharing the DCA data for the phase boundary shown in Fig.~\ref{fig:dop}. Additionally, E.A.S. would like to thank Sergey Brener for valuable discussions. The work of E.A.S. was supported by the Russian Science Foundation, Grant 17-72-20041. The work of M.I.K. was supported by NWO via Spinoza Prize and by ERC Advanced Grant 338957 FEMTO/NANO. A.H. and A.I.L. acknowledge support from the excellence  cluster ``The Hamburg Centre for Ultrafast Imaging - Structure, Dynamics and Control of Matter at the Atomic Scale'' and North-German Supercomputing Alliance (HLRN) under the Project No. hhp00042. Also, the work was partially supported by the Stichting voor Fundamenteel Onderzoek der Materie (FOM), which is financially supported by the Nederlandse Organisatie voor Wetenschappelijk Onderzoek (NWO).
\end{acknowledgments}

\appendix

\section{Bosonic action for the extended Hubbard model}
\label{App1}

Here we explicitly derive a bosonic problem for charge degrees of freedom of the extended Hubbard model. For this reason, one can divide the lattice action~\eqref{eq:actionlatt} into the local impurity ${\cal S}_{\rm imp}$ and nonlocal ${\cal S}_{\rm rem}$ parts following the standard procedure of the DB theory~\cite {Rubtsov20121320, PhysRevB.90.235135}
\begin{align}
\label{eq:actionimpapp}
{\cal S}_{\rm imp} = 
&-\sum\limits_{\nu,\sigma}c^{*}_{\nu\sigma} \left[i\nu+\mu-\Delta^{\phantom{*}}_{\nu}\right] c^{\phantom{*}}_{\nu\sigma} \\
&+U\sum_{\omega}n^{*}_{\omega\uparrow} n^{\phantom{*}}_{\omega\downarrow} 
+\frac12\sum_{\omega}\Lambda^{\phantom{*}}_{\omega}\,\rho^{*}_{\omega}\,\rho^{\phantom{*}}_{\omega}, \notag\\
\label{eq:Simp_SM}
{\cal S}_{\rm rem} = 
&-\sum\limits_{\kv,\nu,\sigma}c^{*}_{\kv\nu\sigma}
\left[\Delta^{\phantom{*}}_{\nu}-\varepsilon^{\phantom{*}}_{\kv}\right]
c^{\phantom{*}}_{\kv\nu\sigma} \\
&+\frac12\sum_{\qv,\omega}
\left[V^{\phantom{*}}_{\qv}-\Lambda^{\phantom{*}}_{\omega}\right]\rho^{*}_{\qv\omega}\,\rho^{\phantom{*}}_{\qv\omega} + \sum\limits_{\qv,\omega} j^{*}_{\qv\omega}\,\rho^{\phantom{*}}_{\qv\omega}, \notag
\end{align}
where we introduced fermionic $\Delta_{\nu}$ and bosonic $\Lambda_{\omega}$ hybridization functions, and sources $j_{\qv\omega}$ for bosonic variables. The partition function of our problem is given by the following relation
\begin{align}
{\cal Z}=\int D[c^{*},c] \, e^{-{\cal S}},
\end{align}
where the action ${\cal S}$ is given by the Eq.~\ref{eq:actionlatt}.
Using the Hubbard--Stratonovich transformation of the reminder term ${\cal S}_{\rm rem}$, one can introduce ${\it dual}$ fermionic $f^{*},f$, and bosonic variables $\phi$ as follows
\begin{align}
&e^{\sum_{\kv,\nu,\sigma} c^{*}_{\kv\nu\sigma}[\Delta^{\phantom{*}}_{\nu}-\varepsilon^{\phantom{*}}_{\kv}]c^{\phantom{*}}_{\kv\nu\sigma}} = \\
&D_{f} \int D[f] \,e^{-\sum_{\kv,\nu,\sigma}\left( f^{*}_{\kv\nu\sigma}[\Delta^{\phantom{*}}_{\nu}-\varepsilon^{\phantom{*}}_{\kv}]^{-1}f^{\phantom{*}}_{\kv\nu\sigma} + c^{*}_{\kv\nu\sigma}f^{\phantom{*}}_{\kv\nu\sigma} + f^{*}_{\kv\nu\sigma}c^{\phantom{*}}_{\kv\nu\sigma}\right)}, \notag\\
&e^{\frac12\sum_{\qv,\omega} \rho^{*}_{\qv\omega}\left[\Lambda^{\phantom{*}}_{\omega}-V^{\phantom{*}}_{\qv}\right]\rho^{\phantom{*}}_{\qv\omega}} = \\
&D_{\phi}\int D[\phi] \,e^{-\left(\frac12\sum_{\qv,\omega} \phi^{*}_{\qv\omega}\left[\Lambda^{\phantom{*}}_{\omega}-V^{\phantom{*}}_{\qv}\right]^{-1}\phi^{\phantom{*}}_{\qv\omega} + \phi^{*}_{\qv\omega}\,\rho^{\phantom{*}}_{\qv\omega}\right)}, \notag
\end{align}
where terms $D_{f} = {\rm det}[\Delta_{\nu}-\varepsilon_{\bf k}]$ and $D^{-1}_{\,\phi} = \sqrt{{\rm det}[\Lambda_{\omega}-V_{\bf q}]}$ can be neglected when calculating expectation values.
Rescaling fermionic and bosonic fields on corresponding Green's functions of the impurity problem as $f^{(*)}_{\kv\nu}\to{}f^{(*)}_{\kv\nu}g^{-1}_{\nu}$ and $\phi_{\qv\omega}\to{}\phi_{\qv\omega}\,\chi^{-1}_{\omega}$, and shifting bosonic variables, the nonlocal part of the action transforms to
\begin{align} 
&{\cal S}_{\rm DB}
= -\sum_{\kv,\nu,\sigma} f^{*}_{\kv\nu\sigma}g^{-1}_{\nu\sigma} [\varepsilon^{\phantom{*}}_{\kv}-\Delta^{\phantom{*}}_{\nu}]^{-1} g^{-1}_{\nu\sigma}f^{\phantom{*}}_{\kv\nu\sigma} \\
&+\sum_{\kv,\nu,\sigma} \left[ c^{*}_{\kv\nu\sigma} g^{-1}_{\nu\sigma}f^{\phantom{*}}_{\kv\nu\sigma} + f^{*}_{\kv\nu\sigma} g^{-1}_{\nu\sigma}c^{\phantom{*}}_{\kv\nu\sigma} \right] + \sum_{\qv,\omega} \phi^{\phantom{*}}_{\qv\omega}\,\chi^{-1}_{\omega}\rho^{\phantom{*}}_{\qv\omega}
\notag\\
&-\frac12\sum_{\qv,\omega} \left( \phi^{*}_{\qv\omega} - j^{*}_{\qv\omega} \,\chi^{\phantom{*}}_{\omega}\right)\chi^{-1}_{\omega} 
\left[V^{\phantom{*}}_{\qv} - \Lambda^{\phantom{*}}_{\omega}\right]^{-1} \chi^{-1}_{\omega}\left( \phi^{\phantom{*}}_{\qv\omega} - \chi^{\phantom{*}}_{\omega}j^{\phantom{*}}_{\qv\omega}\right). \notag
\end{align}
Integrating out initial degrees of freedom with respect to the impurity action~\eqref{eq:actionimpapp}, one gets~\cite{Rubtsov20121320} 
\begin{align}
\int &D[c]\,e^{-\sum_{i}{\cal S}^{\,(i)}_{\rm imp} - \sum_{\kv,\nu,\sigma} \left[ c^{*}_{\kv\nu\sigma} g^{-1}_{\nu\sigma}f^{\phantom{*}}_{\kv\nu\sigma} + 
f^{*}_{\kv\nu\sigma} g^{-1}_{\nu\sigma}c^{\phantom{*}}_{\kv\nu\sigma} \right] 
-\sum_{\qv,\omega} \phi^{*}_{\qv\omega}\, \chi^{-1}_{\omega} \rho^{\phantom{*}}_{\qv\omega}} = \notag\\
&{\cal Z}_{\rm imp}\times e^{ -\sum_{\kv,\nu,\sigma}f^{*}_{\kv\nu\sigma} g^{-1}_{\nu\sigma}f^{\phantom{*}}_{\kv\nu\sigma} 
-\frac12\sum_{\qv,\omega} \phi^{*}_{\qv\omega} \,\chi^{-1}_{\omega}\phi^{\phantom{*}}_{\qv\omega} - \tilde{W}[f,\phi]},
\end{align}
where ${\cal Z}_{\rm imp}$ is a partition function of the impurity problem. Here, the interaction $\tilde{W}[f,\phi]$ is presented as an infinite series of full vertex functions of the local impurity problem~\eqref{eq:actionimpapp} ~\cite{Rubtsov20121320, PhysRevB.93.045107}. The lowest order interaction terms are 
\begin{align}
\label{eq:lowestint}
\tilde{W}[f,\phi]
&=\sum_{\kv,\kv',\qv}\sum_{\nu,\nu',\omega}\sum_{\sigma(')}\left( 
\phi^{*}_{\qv\omega}\gamma^{\phantom{*}}_{\nu\omega}\,f^{*}_{\kv\nu\sigma} f^{\phantom{*}}_{\kv+\qv,\nu+\omega,\sigma} \right.\\
&\left. - \frac14\,\overline{\gamma}^{\phantom{*}}_{\nu\nu'\omega}\, f^{*}_{\kv\nu\sigma}f^{\phantom{*}}_{\kv+\qv,\nu+\omega,\sigma'}f^{*}_{\kv'+\qv,\nu'+\omega,\sigma''} f^{\phantom{*}}_{\kv'\nu'\sigma'''}\right), \notag
\end{align}
where the full three- and four-point vertex functions are defined as
\begin{align}
\label{eq:4vertapp}
\gamma^{\phantom{*}}_{\nu\omega}  
&= \av{c^{\phantom{*}}_{\nu\sigma}\,c^{*}_{\nu+\omega,\sigma}\,\rho^{\phantom{*}}_{\omega}}_{\rm imp} \,\chi_{\omega}^{-1}g^{-1}_{\nu\sigma}\,g^{-1}_{\nu+\omega,\sigma},\\
\overline{\gamma}_{\nu\nu'\omega} &=
\av{c^{\phantom{*}}_{\nu\sigma} c^{*}_{\nu+\omega,\sigma'} c^{\phantom{*}}_{\nu'+\omega,\sigma''} c^{*}_{\nu'\sigma'''}}_{\rm c~ imp}\,g^{-1}_{\nu\sigma}\,g^{-1}_{\nu+\omega,\sigma'}\,g^{-1}_{\nu'+\omega,\sigma''}\, g^{-1}_{\nu'\sigma'''}. \notag
\end{align}
Note that the four-point vertex $\overline{\gamma}_{\nu\nu'\omega}$ is defined here in the particle-hole channel.

Therefore, the initial lattice problem transforms to the following {\it dual} action 
\begin{align}
&{\cal \tilde{S}}
= -\sum_{\kv,\nu,\sigma} f^{*}_{\kv\nu\sigma}g^{-1}_{\nu\sigma} [\varepsilon^{\phantom{*}}_{\kv}-\Delta^{\phantom{*}}_{\nu}]^{-1} g^{-1}_{\nu\sigma}f^{\phantom{*}}_{\kv\nu\sigma} \\
&+\sum_{\kv,\nu}f^{*}_{\kv\nu\sigma}\, g^{-1}_{\nu\sigma}f^{\phantom{*}}_{\kv\nu\sigma} 
+\frac12\sum_{\qv,\omega} \phi^{*}_{\qv\omega}\, \chi^{-1}_{\omega}\phi^{\phantom{*}}_{\qv\omega} + \tilde{W}[f,\phi] \notag \\
&-\frac12\sum_{\qv,\omega} \left( \phi^{*}_{\qv\omega} - j^{*}_{\qv\omega}\, \chi^{\phantom{*}}_{\omega} \right)\chi^{-1}_{\omega} 
\left[V^{\phantom{*}}_{\qv} - \Lambda^{\phantom{*}}_{\omega}\right]^{-1} \chi^{-1}_{\omega} \left( \phi^{\phantom{*}}_{\qv\omega} - \chi^{\phantom{*}}_{\omega} j^{\phantom{*}}_{\qv\omega}\right) .\notag
\end{align}
In order to come back to the original bosonic variables, one can perform the third Hubbard-Stratonovich transformation as 
\begin{align}
&e^{\frac12\sum_{\qv,\omega} \left( \phi^{*}_{\qv\omega} - j^{*}_{\qv\omega}\, \chi^{\phantom{*}}_{\omega} \right)
\chi^{-1}_{\omega} \left[V^{\phantom{*}}_{\qv} - \Lambda^{\phantom{*}}_{\omega}\right]^{-1} \chi^{-1}_{\omega} \left(\phi^{\phantom{*}}_{\qv\omega} - \chi^{\phantom{*}}_{\omega}
j^{\phantom{*}}_{\qv\omega}\right)} = \\
&D_{\bar\rho}
\int D[\bar\rho]\,e^{-\sum_{\qv,\omega}\,\left( \frac12T\bar\rho_{\qv\omega}\,
\left[V_{\qv} - \Lambda_{\omega} \right]\bar\rho_{\qv\omega} - \phi_{\qv\omega}\, \chi^{-1}_{\omega} \bar\rho_{-\qv,-\omega} + j_{\qv\omega}\, \bar\rho_{-\qv,-\omega} \right)}. \notag
\end{align}
Comparing this expression to the Eq.~\ref{eq:Simp_SM}, one can see that sources $j^{*}_{\qv\omega}$ introduced for the initial degrees of freedom $\rho_{\qv\omega}$ are also the sources for new bosonic fields $\bar\rho_{\qv\omega}$. Therefore, fields $\bar\rho_{\qv\omega}$ indeed represent initial degrees of freedom and have the same physical meaning as original {\it composite} bosonic variables $\rho_{\qv\omega}=\sum_{\kv\nu\sigma}c^{*}_{\kv\nu\sigma} c^{\phantom{*}}_{\kv+\qv,\nu+\omega,\sigma}-\av{n^{\phantom{*}}_{\qv\omega}}$ of the lattice problem~\eqref{eq:actionlatt}. Nevertheless, $\bar\rho_{\qv\omega}$ can now be treated as {\it elementary} bosonic fields that have a well-defined propagator, since they are introduced as a decoupling fields of dual degrees of freedom $\phi_{\qv\omega}$ and therefore, independent on fermionic variables $c^{*}_{\kv\nu\sigma}\,(c^{\phantom{*}}_{\kv\nu\sigma})$. Taking sources to zero and replacing $\bar\rho_{\qv\omega}$ by $\rho_{\qv\omega}$, dual bosonic fields can be integrated out as~\cite{PhysRevLett.121.037204}
\begin{align}
\label{eq:integrationphi}
\int &D[\phi]\,e^{-\frac12\sum_{\qv,\omega} \phi^{*}_{\qv\omega} \,\chi^{-1}_{\omega}\phi^{\phantom{*}}_{\qv\omega} - \phi^{*}_{\qv\omega}\, \chi^{-1}_{\omega} \rho^{\phantom{*}}_{\qv\omega} - \tilde{W}[f,\phi]}
= \\
&{\cal Z}_{\phi}\times e^{\frac12\sum_{\qv,\omega} \rho^{*}_{\qv\omega}\, \chi_{\omega}^{-1} \rho^{\phantom{*}}_{\qv\omega} - W[f,\rho]}, \notag
\end{align}
where ${\cal Z}_{\phi}$ is a partition function of the Gaussian part of the bosonic action. Here, we restrict ourselves to the lowest order interaction terms of $\tilde{W}[f,\phi]$ shown in Eq.~\ref{eq:lowestint}. Then, the integration of dual bosonic fields in Eq.~\ref{eq:integrationphi} simplifies and $W[f,\rho]$ keeps an efficient dual form of $\tilde{W}[f,\phi]$~\eqref{eq:lowestint} with replacement of bosonic variables $\phi\to\rho$ 
\begin{align}
\label{eq:Wfull}
&W[f,\rho]
=\sum_{\kv,\kv',\qv}\sum_{\nu,\nu',\omega}\sum_{\sigma(')}\left(\rho^{*}_{\qv\omega} 
\gamma^{\phantom{*}}_{\nu\omega} f^{*}_{\kv\nu\sigma} f^{\phantom{*}}_{\kv+\qv,\nu+\omega,\sigma} \right.\\
&\left.- \left[\overline{\gamma}_{\nu\nu'\omega} + \gamma^{\phantom{*}}_{\nu\omega}\,\chi^{\phantom{*}}_{\omega}\gamma^{\phantom{*}}_{\nu'+\omega,-\omega}\right] f^{*}_{\kv\nu\sigma}f^{\phantom{*}}_{\kv+\qv,\nu+\omega,\sigma}f^{*}_{\kv'+\qv,\nu'+\omega,\sigma'} f^{\phantom{*}}_{\kv'\nu'\sigma'}\right). \notag
\end{align}
As can be seen in Ref.~\onlinecite{PhysRevLett.121.037204}, the four-point vertex becomes irreducible with respect to the full local bosonic propagator $\chi_{\omega}$, while the three-point vertex $\gamma_{\nu\omega}$ remains invariant.
Therefore, the problem transforms to the following action of an effective $s$-$d$ model 
\begin{align}
{\cal S}_{s\text{-}d} = &-\sum_{\kv,\nu,\sigma} f^{*}_{{\bf k}\nu\sigma}\tilde{G}^{-1}_{0}f^{\phantom{*}}_{{\bf k}\nu\sigma} 
-\frac12\sum_{\qv,\omega} \rho^{*}_{\qv\omega}X^{-1}_{\rm E}\rho^{\phantom{*}}_{\qv\omega} + W[f,\rho],
\label{eq:actionSDapp}
\end{align}
where 
\begin{align}
X_{\rm E} = \left[\chi^{-1}_{\omega} + \Lambda^{\phantom{1}}_{\omega} - V^{\phantom{1}}_{\qv} \right]^{-1}
\end{align}
is the EDMFT susceptibility and $\tilde{G}_{0}$ is a nonlocal part of the EDMFT Green's function. When the main contribution to the four-point vertex is given by the reducible contribution with respect to the full local bosonic propagator, i.e. 
\begin{align}
\label{eq:PiDualApp}
\overline{\gamma}_{\nu\nu'\omega}\simeq-\gamma_{\nu\omega}\,\chi_{\omega}\,\gamma_{\nu'+\omega,-\omega}= \vcenter{\hbox{\includegraphics[width=0.15\linewidth]{ChiVert.pdf}}}\,,
\end{align}
the interaction part of the action~\eqref{eq:actionSDapp} takes the most simple form that contains only the three-point vertex function
\begin{align}
W'[f,\rho]
\simeq\sum_{\kv,\qv}\sum_{\nu,\omega,\sigma}
\rho^{*}_{\qv\omega} \gamma^{\phantom{*}}_{\nu\omega} f^{*}_{\kv\nu\sigma} f^{\phantom{*}}_{\kv+\qv,\nu+\omega,\sigma}.
\label{eq:Wsimple}
\end{align}

\begin{figure*}[t!]
\begin{center}
\includegraphics[width=0.24\linewidth]{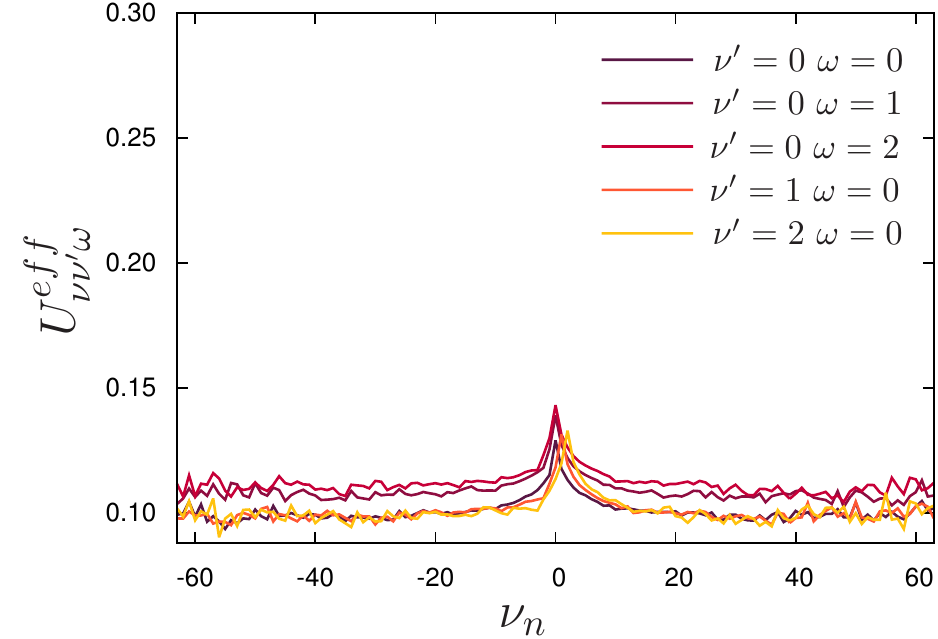}~
\includegraphics[width=0.24\linewidth]{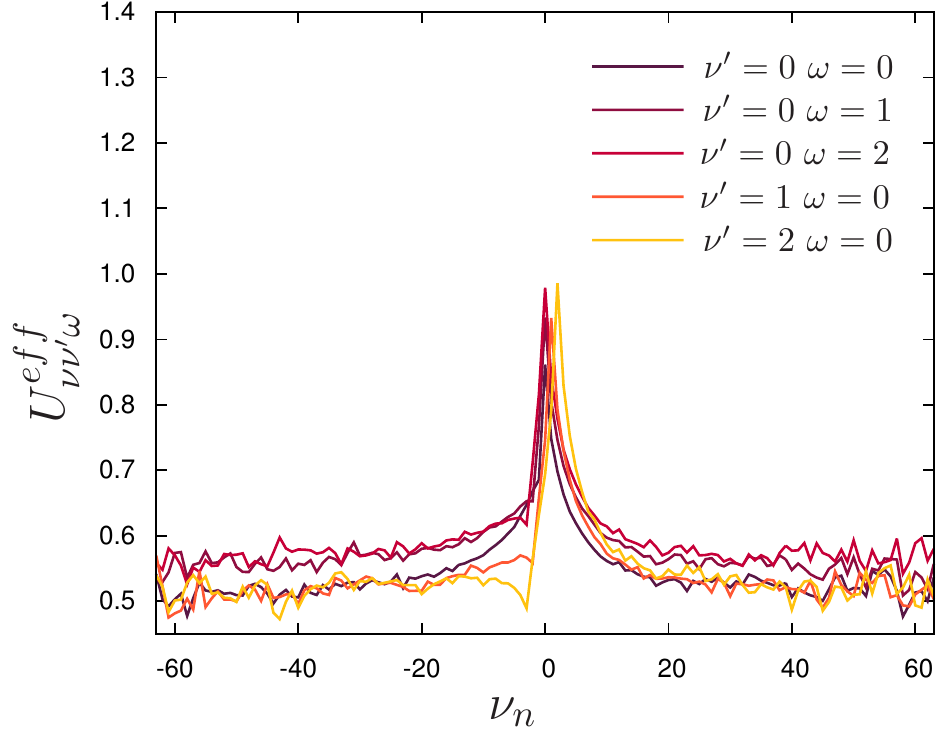}~
\includegraphics[width=0.24\linewidth]{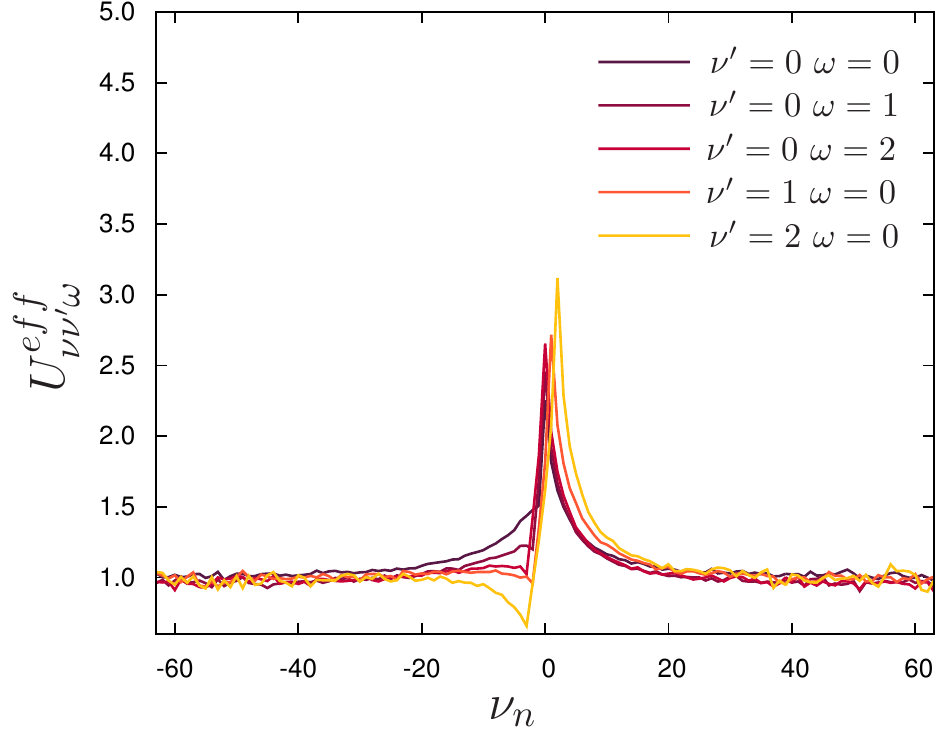}~
\includegraphics[width=0.24\linewidth]{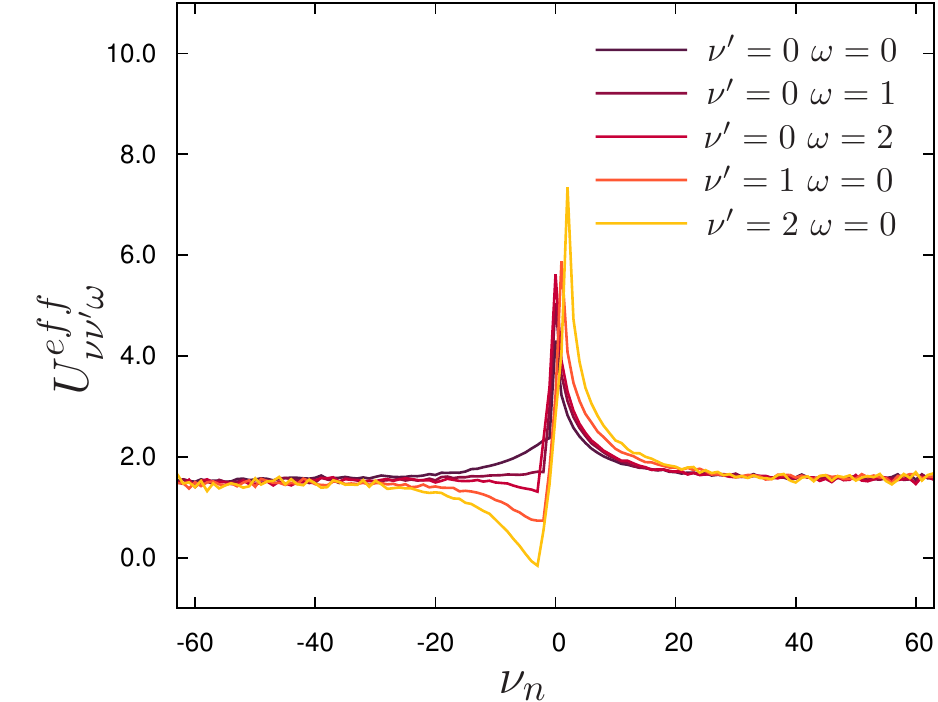}
\end{center}
\vspace{-0.3cm}
\caption{Frequency dependence of the effective local Coulomb interaction $U^{\rm eff}_{\nu\nu'\omega}$ obtained for different values of the actual Coulomb interaction $U=0.1;\,0.5;\,1.0;\,1.5$ (from left to right) in the normal phase close to the charge ordering for different values of fermionic $\nu'$ and bosonic $\omega$ frequencies for $\beta=50$. The dependence of the effective interaction on fermionic frequency becomes larger at larger Coulomb interaction. \label{fig:SM1}}
\end{figure*}

According to derivations presented in Ref.~\onlinecite{PhysRevLett.121.037204}, one can integrate out {\it dual} fermionic degrees of freedom using the ladder approximation and obtain an effective problem written in terms of bosonic degrees of freedom only
\begin{align}
{\cal S}
= -\frac12\sum_{\qv,\omega} \rho^{*}_{\qv\omega}
X^{-1}_{\qv\omega}\,\rho^{\phantom{*}}_{\qv\omega}, 
\label{eq:spinactionapp}
\end{align}
where the expression for the lattice susceptibility reads
\begin{align}
\left[X^{\rm ladd}_{\qv\omega}\right]^{-1} = \left[X^{\rm DMFT}_{\qv\omega}\right]^{-1} + \Lambda_{\omega} - V_{\qv}.
\label{eq:X_DBapp}
\end{align}
Here, 
\begin{align}
\hat{X}^{\rm DMFT}_{\qv\omega} = \Tr\left\{\hat{X}^{0}_{\qv\omega} \left[I + \hat{\overline{\gamma}}^{\,\rm 2PI}_{\omega}  \hat{X}^{0}_{\qv\omega} \right]^{-1}\right\}
\end{align}
is the DMFT-like~\cite{PhysRevLett.62.324, RevModPhys.68.13}
susceptibility written in terms of lattice Green's functions, and 2PI vertex functions of impurity model defined as
\begin{align}
\hat{\overline{\gamma}}^{\rm 2PI}_{\omega} =  \hat{\overline{\gamma}}^{\phantom{2}}_{\omega}\left[I-\hat{\chi}^{0}_{\omega}\,\hat{\overline{\gamma}}^{\phantom{2}}_{\omega}\right]^{-1}.
\end{align}
Here, multiplication and inversion should be understood as a standard matrix operations in the space of fermionic frequencies $\nu,\nu'$. $I$ is the identity matrix in the same space, and the trace is taken over the external fermionic indices. For simplicity, we omit fermionic indices wherever they are not crucial for understanding. Matrix elements of the bare lattice $\hat{X}^{0}_{\qv\omega}$ and local impurity $\hat{\chi}^{0}_{\omega}$ charge susceptibilities are defined as 
\begin{align}
X^{0}_{\qv\omega;\,\nu\nu'} &= \sum_{\kv,\sigma}G_{\kv+\qv,\nu+\omega,\sigma}G_{\kv\nu\sigma}\,\delta_{\nu\nu'}\\
\chi^{0}_{\omega;\,\nu\nu'} &= \sum_{\sigma}g_{\nu+\omega,\sigma}g_{\nu\sigma}\delta_{\nu\nu'}.
\end{align} 
Matrix elements $\overline{\gamma}_{\nu\nu'\omega}$ of the four-point vertex function $\hat{\overline{\gamma}}_{\omega}$ are defined in ~\eqref{eq:4vertapp}.

Therefore, the charge susceptibility~\eqref{eq:X_DBapp} in the ladder approximation can be rewritten as
\begin{align}
X^{\rm ladd}_{\qv\omega} = \Tr\left\{\hat{X}^{0}_{\qv\omega} \left[ I - \left(I\,V^{\phantom{*}}_{\bf q} + \hat{U}^{\rm eff}_{\omega}\right) \hat{X}^{0}_{\qv\omega} \right]^{-1}\right\},
\label{eq:Xladdapp}
\end{align}
where we introduced an effective bare local interaction
\begin{align}
U^{\rm eff}_{\nu\nu'\omega} = -\Lambda^{\phantom{2}}_{\omega} - \overline{\gamma}^{\,\rm 2PI}_{\nu\nu'\omega}
\end{align}
shown in Fig.~\ref{fig:SM1} for different values of fermionic $\nu'$ and bosonic $\omega$ frequencies.

Another simplified expression for the charge susceptibility can be obtained after expanding the simplified form of interaction $W[f,\rho]$ given by Eq.~\ref{eq:Wsimple} up to the second order with respect to bosonic fields $\rho$ in the expression for the partition function of the action~\eqref{eq:actionSDapp}. This results in 
\begin{align}
\left[X^{(2)}_{\qv\omega}\right]^{-1}
&= -V^{\phantom{1}}_{\bf q} + \Lambda^{\phantom{1}}_{\omega} + \chi^{-1}_{\omega} - \tilde{\Pi}^{(2)}_{\qv\omega},
\label{eq:X1pmapp}
\end{align}
where
\begin{align}
\tilde{\Pi}^{(2)}_{\qv\omega}=\sum_{\kv,\nu,\sigma}
\gamma^{\phantom{1}}_{\nu+\omega,-\omega}\,\tilde{G}^{\phantom{2}}_{\kv+\qv,\nu+\omega\sigma}\tilde{G}^{\phantom{2}}_{\kv\nu\sigma}\,\gamma^{\phantom{1}}_{\nu,\omega}
\end{align}
is the second order polarization operator and $\tilde{G}_{\kv\nu}$ is a nonlocal part of the lattice (EDMFT) Green's function. As discussed in the main text, this expression can be transformed to a pair interaction of the classical Ising model. 

\section{vertex approximation}
\label{App3}

According to discussions presented in the main text, the expression for the 2PI four-point vertex function can be approximated as $\overline{\gamma}^{\rm 2PI}_{\nu\nu'\omega}\simeq\overline{\gamma}^{\rm 2PI}_{\omega}$ when its dependence on fermionic frequencies is negligible. Then, using the exact relation for the local impurity susceptibility
\begin{align}
\chi_{\omega} = \Tr\left\{\hat{\chi}^{0}_{\omega} - \hat{\chi}^{0}_{\omega}\,\hat{\overline{\gamma}}_{\omega} \,\hat{\chi}^{0}_{\omega} \right\} = \Tr\left\{\hat{\chi}^{0}_{\omega} \left[I+\hat{\overline{\gamma}}^{\rm 2PI}_{\omega}\,\hat{\chi}^{0}_{\omega}\right]^{-1}\right\}
\end{align} 
and assuming that the 2PI vertex does not depend on fermionic frequencies, one gets
\begin{align}
\overline\gamma^{\rm 2PI}_{\nu\nu'\omega}\simeq\overline\gamma^{\rm 2PI}_{\omega}=\chi^{-1}_{\omega}-\chi^{0~-1}_{\omega}.
\label{eq:gamma2PIapp}
\end{align}
As shown in Ref.~\onlinecite{PhysRevLett.121.037204}, in the case of well-developed collective fluctuations the four-point function is described by the bosonic frequency and three-point vertex function that enters the exact Hedin equation~\cite{PhysRev.139.A796} for the self-energy and polarization operator of the impurity problem is close to unity. As a consequence, the local self-energy and polarization operator take the same form as in GW approach~\cite{PhysRev.139.A796, 0034-4885-61-3-002, 0953-8984-11-42-201}. Thus, the polarization operator of the impurity problem can be approximated as $\Pi_{\omega} \simeq \chi^{0}_{\omega}$ neglecting the vertex function. Using the exact relation for the local charge susceptibility of the impurity problem, one gets the following relation
\begin{align}
\chi^{-1}_{\omega} = \Pi^{-1}_{\omega} - \left(U + \Lambda^{\phantom{1}}_{\omega}\right) \simeq \chi^{0~-1}_{\omega} - \left(U + \Lambda^{\phantom{1}}_{\omega}\right).
\end{align}
Therefore, in the regime of strong charge fluctuations the 2PI vertex function can be approximated as
$
\overline\gamma^{\rm 2PI}_{\nu\nu'\omega}\simeq-U-\Lambda^{\phantom{1}}_{\omega}.
$

The three-point vertex can also be approximated using the exact relation between three- and four-point vertex functions, and the simplified form of the 2PI vertex~\cite{PhysRevLett.121.037204} 
\begin{align}
\label{eq:3approxWapp}
\gamma_{\nu\omega} \simeq \gamma_{\omega}
= \chi^{-1}_{\omega} -\overline\gamma^{\rm 2PI}_{\omega} = \chi^{-1}_{\omega} + \Lambda^{\phantom{1}}_{\omega} + U^{\rm eff}_{\omega},
\end{align}
where $U^{\rm eff}_{\omega} = -\Lambda^{\phantom{1}}_{\omega} - \overline\gamma^{\rm 2PI}_{\omega}$.
Taking into account that in the regime of well-developed charge fluctuations the effective interaction coincides with the actual value of the bare local Coulomb interaction $U^{\rm eff}_{\omega}\simeq{}U$, one can further write
\begin{align}
\gamma_{\omega} 
&\simeq \chi^{-1}_{\omega} + \Lambda^{\phantom{1}}_{\omega} + U = \Pi^{-1}_{\omega} = \frac{U{\cal W}_{\omega}}{{\cal W}_{\omega}-U} = \frac{U}{1-\varepsilon_{\omega}}, 
\end{align}
where we introduced the renormalized local Coulomb interaction 
$
{\cal W}_{\omega} = U/(1-\Pi_{\omega}U)
$
that is connected to the bare Coulomb interaction via the dielectric function $\varepsilon_{\omega}=U/{\cal W}_{\omega}$.

\bibliography{Charge_GW}

\begin{thebibliography}{66}%
\makeatletter
\providecommand \@ifxundefined [1]{%
 \@ifx{#1\undefined}
}%
\providecommand \@ifnum [1]{%
 \ifnum #1\expandafter \@firstoftwo
 \else \expandafter \@secondoftwo
 \fi
}%
\providecommand \@ifx [1]{%
 \ifx #1\expandafter \@firstoftwo
 \else \expandafter \@secondoftwo
 \fi
}%
\providecommand \natexlab [1]{#1}%
\providecommand \enquote  [1]{``#1''}%
\providecommand \bibnamefont  [1]{#1}%
\providecommand \bibfnamefont [1]{#1}%
\providecommand \citenamefont [1]{#1}%
\providecommand \href@noop [0]{\@secondoftwo}%
\providecommand \href [0]{\begingroup \@sanitize@url \@href}%
\providecommand \@href[1]{\@@startlink{#1}\@@href}%
\providecommand \@@href[1]{\endgroup#1\@@endlink}%
\providecommand \@sanitize@url [0]{\catcode `\\12\catcode `\$12\catcode
  `\&12\catcode `\#12\catcode `\^12\catcode `\_12\catcode `\%12\relax}%
\providecommand \@@startlink[1]{}%
\providecommand \@@endlink[0]{}%
\providecommand \url  [0]{\begingroup\@sanitize@url \@url }%
\providecommand \@url [1]{\endgroup\@href {#1}{\urlprefix }}%
\providecommand \urlprefix  [0]{URL }%
\providecommand \Eprint [0]{\href }%
\providecommand \doibase [0]{http://dx.doi.org/}%
\providecommand \selectlanguage [0]{\@gobble}%
\providecommand \bibinfo  [0]{\@secondoftwo}%
\providecommand \bibfield  [0]{\@secondoftwo}%
\providecommand \translation [1]{[#1]}%
\providecommand \BibitemOpen [0]{}%
\providecommand \bibitemStop [0]{}%
\providecommand \bibitemNoStop [0]{.\EOS\space}%
\providecommand \EOS [0]{\spacefactor3000\relax}%
\providecommand \BibitemShut  [1]{\csname bibitem#1\endcsname}%
\let\auto@bib@innerbib\@empty
\bibitem [{\citenamefont {Pines}\ and\ \citenamefont
  {Nozieres}(1966)}]{pines1966theory}%
  \BibitemOpen
  \bibfield  {author} {\bibinfo {author} {\bibfnamefont {D.}~\bibnamefont
  {Pines}}\ and\ \bibinfo {author} {\bibfnamefont {P.}~\bibnamefont
  {Nozieres}},\ }\href@noop {} {\emph {\bibinfo {title} {{The Theory of Quantum
  Liquids: Normal Fermi liquids. Theory of quantum liquids}}}}\ (\bibinfo
  {publisher} {W.A. Benjamin, Philadelphia},\ \bibinfo {year}
  {1966})\BibitemShut {NoStop}%
\bibitem [{\citenamefont {Platzman}\ and\ \citenamefont
  {Wolff}(1973)}]{platzman1973waves}%
  \BibitemOpen
  \bibfield  {author} {\bibinfo {author} {\bibfnamefont {P.~M.}\ \bibnamefont
  {Platzman}}\ and\ \bibinfo {author} {\bibfnamefont {P.~A.}\ \bibnamefont
  {Wolff}},\ }\href@noop {} {\emph {\bibinfo {title} {{Waves and interactions
  in solid state plasmas}}}},\ Vol.~\bibinfo {volume} {13}\ (\bibinfo
  {publisher} {Academic Press, New York},\ \bibinfo {year} {1973})\BibitemShut
  {NoStop}%
\bibitem [{\citenamefont {Vonsovsky}\ and\ \citenamefont
  {Katsnelson}(1989)}]{vonsovsky1989quantum}%
  \BibitemOpen
  \bibfield  {author} {\bibinfo {author} {\bibfnamefont {S.~V.}\ \bibnamefont
  {Vonsovsky}}\ and\ \bibinfo {author} {\bibfnamefont {M.~I.}\ \bibnamefont
  {Katsnelson}},\ }\href@noop {} {\emph {\bibinfo {title} {{Quantum solid-state
  physics}}}}\ (\bibinfo  {publisher} {Springer Verlag, Berlin},\ \bibinfo
  {year} {1989})\BibitemShut {NoStop}%
\bibitem [{\citenamefont {Ayral}\ \emph {et~al.}(2013)\citenamefont {Ayral},
  \citenamefont {Biermann},\ and\ \citenamefont {Werner}}]{PhysRevB.87.125149}%
  \BibitemOpen
  \bibfield  {author} {\bibinfo {author} {\bibfnamefont {T.}~\bibnamefont
  {Ayral}}, \bibinfo {author} {\bibfnamefont {S.}~\bibnamefont {Biermann}}, \
  and\ \bibinfo {author} {\bibfnamefont {P.}~\bibnamefont {Werner}},\
  }\bibfield  {title} {\enquote {\bibinfo {title} {{Screening and nonlocal
  correlations in the extended Hubbard model from self-consistent combined $GW$
  and dynamical mean field theory}},}\ }\href {\doibase
  10.1103/PhysRevB.87.125149} {\bibfield  {journal} {\bibinfo  {journal} {Phys.
  Rev. B}\ }\textbf {\bibinfo {volume} {87}},\ \bibinfo {pages} {125149}
  (\bibinfo {year} {2013})}\BibitemShut {NoStop}%
\bibitem [{\citenamefont {Hafermann}\ \emph {et~al.}(2014)\citenamefont
  {Hafermann}, \citenamefont {van Loon}, \citenamefont {Katsnelson},
  \citenamefont {Lichtenstein},\ and\ \citenamefont
  {Parcollet}}]{PhysRevB.90.235105}%
  \BibitemOpen
  \bibfield  {author} {\bibinfo {author} {\bibfnamefont {H.}~\bibnamefont
  {Hafermann}}, \bibinfo {author} {\bibfnamefont {E.~G. C.~P.}\ \bibnamefont
  {van Loon}}, \bibinfo {author} {\bibfnamefont {M.~I.}\ \bibnamefont
  {Katsnelson}}, \bibinfo {author} {\bibfnamefont {A.~I.}\ \bibnamefont
  {Lichtenstein}}, \ and\ \bibinfo {author} {\bibfnamefont {O.}~\bibnamefont
  {Parcollet}},\ }\bibfield  {title} {\enquote {\bibinfo {title} {{Collective
  charge excitations of strongly correlated electrons, vertex corrections, and
  gauge invariance}},}\ }\href {\doibase 10.1103/PhysRevB.90.235105} {\bibfield
   {journal} {\bibinfo  {journal} {Phys. Rev. B}\ }\textbf {\bibinfo {volume}
  {90}},\ \bibinfo {pages} {235105} (\bibinfo {year} {2014})}\BibitemShut
  {NoStop}%
\bibitem [{\citenamefont {van Loon}\ \emph
  {et~al.}(2014{\natexlab{a}})\citenamefont {van Loon}, \citenamefont
  {Hafermann}, \citenamefont {Lichtenstein}, \citenamefont {Rubtsov},\ and\
  \citenamefont {Katsnelson}}]{PhysRevLett.113.246407}%
  \BibitemOpen
  \bibfield  {author} {\bibinfo {author} {\bibfnamefont {E.~G. C.~P.}\
  \bibnamefont {van Loon}}, \bibinfo {author} {\bibfnamefont {H.}~\bibnamefont
  {Hafermann}}, \bibinfo {author} {\bibfnamefont {A.~I.}\ \bibnamefont
  {Lichtenstein}}, \bibinfo {author} {\bibfnamefont {A.~N.}\ \bibnamefont
  {Rubtsov}}, \ and\ \bibinfo {author} {\bibfnamefont {M.~I.}\ \bibnamefont
  {Katsnelson}},\ }\bibfield  {title} {\enquote {\bibinfo {title} {{Plasmons in
  Strongly Correlated Systems: Spectral Weight Transfer and Renormalized
  Dispersion}},}\ }\href {\doibase 10.1103/PhysRevLett.113.246407} {\bibfield
  {journal} {\bibinfo  {journal} {Phys. Rev. Lett.}\ }\textbf {\bibinfo
  {volume} {113}},\ \bibinfo {pages} {246407} (\bibinfo {year}
  {2014}{\natexlab{a}})}\BibitemShut {NoStop}%
\bibitem [{\citenamefont {Rubtsov}\ \emph {et~al.}(2012)\citenamefont
  {Rubtsov}, \citenamefont {Katsnelson},\ and\ \citenamefont
  {Lichtenstein}}]{Rubtsov20121320}%
  \BibitemOpen
  \bibfield  {author} {\bibinfo {author} {\bibfnamefont {A.~N.}\ \bibnamefont
  {Rubtsov}}, \bibinfo {author} {\bibfnamefont {M.~I.}\ \bibnamefont
  {Katsnelson}}, \ and\ \bibinfo {author} {\bibfnamefont {A.~I.}\ \bibnamefont
  {Lichtenstein}},\ }\bibfield  {title} {\enquote {\bibinfo {title} {{Dual
  boson approach to collective excitations in correlated fermionic systems}},}\
  }\href {\doibase http://dx.doi.org/10.1016/j.aop.2012.01.002} {\bibfield
  {journal} {\bibinfo  {journal} {Annals of Physics}\ }\textbf {\bibinfo
  {volume} {327}},\ \bibinfo {pages} {1320 -- 1335} (\bibinfo {year}
  {2012})}\BibitemShut {NoStop}%
\bibitem [{\citenamefont {Stepanov}\ \emph
  {et~al.}(2016{\natexlab{a}})\citenamefont {Stepanov}, \citenamefont {van
  Loon}, \citenamefont {Katanin}, \citenamefont {Lichtenstein}, \citenamefont
  {Katsnelson},\ and\ \citenamefont {Rubtsov}}]{PhysRevB.93.045107}%
  \BibitemOpen
  \bibfield  {author} {\bibinfo {author} {\bibfnamefont {E.~A.}\ \bibnamefont
  {Stepanov}}, \bibinfo {author} {\bibfnamefont {E.~G. C.~P.}\ \bibnamefont
  {van Loon}}, \bibinfo {author} {\bibfnamefont {A.~A.}\ \bibnamefont
  {Katanin}}, \bibinfo {author} {\bibfnamefont {A.~I.}\ \bibnamefont
  {Lichtenstein}}, \bibinfo {author} {\bibfnamefont {M.~I.}\ \bibnamefont
  {Katsnelson}}, \ and\ \bibinfo {author} {\bibfnamefont {A.~N.}\ \bibnamefont
  {Rubtsov}},\ }\bibfield  {title} {\enquote {\bibinfo {title}
  {{Self-consistent dual boson approach to single-particle and collective
  excitations in correlated systems}},}\ }\href {\doibase
  10.1103/PhysRevB.93.045107} {\bibfield  {journal} {\bibinfo  {journal} {Phys.
  Rev. B}\ }\textbf {\bibinfo {volume} {93}},\ \bibinfo {pages} {045107}
  (\bibinfo {year} {2016}{\natexlab{a}})}\BibitemShut {NoStop}%
\bibitem [{\citenamefont {Verwey}\ and\ \citenamefont
  {Haayman}(1941)}]{VERWEY1941979}%
  \BibitemOpen
  \bibfield  {author} {\bibinfo {author} {\bibfnamefont {E.~J.~W.}\
  \bibnamefont {Verwey}}\ and\ \bibinfo {author} {\bibfnamefont {P.~W.}\
  \bibnamefont {Haayman}},\ }\bibfield  {title} {\enquote {\bibinfo {title}
  {{Electronic conductivity and transition point of magnetite
  {Fe$_3$O$_4$}}},}\ }\href {\doibase
  https://doi.org/10.1016/S0031-8914(41)80005-6} {\bibfield  {journal}
  {\bibinfo  {journal} {Physica}\ }\textbf {\bibinfo {volume} {8}},\ \bibinfo
  {pages} {979 -- 987} (\bibinfo {year} {1941})}\BibitemShut {NoStop}%
\bibitem [{\citenamefont {Verwey}\ \emph {et~al.}(1947)\citenamefont {Verwey},
  \citenamefont {Haayman},\ and\ \citenamefont
  {Romeijn}}]{doi:10.1063/1.1746466}%
  \BibitemOpen
  \bibfield  {author} {\bibinfo {author} {\bibfnamefont {E.~J.~W.}\
  \bibnamefont {Verwey}}, \bibinfo {author} {\bibfnamefont {P.~W.}\
  \bibnamefont {Haayman}}, \ and\ \bibinfo {author} {\bibfnamefont {F.~C.}\
  \bibnamefont {Romeijn}},\ }\bibfield  {title} {\enquote {\bibinfo {title}
  {{Physical Properties and Cation Arrangement of Oxides with Spinel Structures
  II. Electronic Conductivity}},}\ }\href {\doibase 10.1063/1.1746466}
  {\bibfield  {journal} {\bibinfo  {journal} {The Journal of Chemical Physics}\
  }\textbf {\bibinfo {volume} {15}},\ \bibinfo {pages} {181--187} (\bibinfo
  {year} {1947})}\BibitemShut {NoStop}%
\bibitem [{\citenamefont {Mott}(1974)}]{Mott}%
  \BibitemOpen
  \bibfield  {author} {\bibinfo {author} {\bibfnamefont {N.~F.}\ \bibnamefont
  {Mott}},\ }\href@noop {} {\emph {\bibinfo {title} {{Metal-insulator
  transitions}}}}\ (\bibinfo  {publisher} {London: Taylor \& Francis},\
  \bibinfo {year} {1974})\BibitemShut {NoStop}%
\bibitem [{\citenamefont {Staub}\ \emph {et~al.}(2005)\citenamefont {Staub},
  \citenamefont {Shi}, \citenamefont {Schulze-Briese}, \citenamefont
  {Patterson}, \citenamefont {Fauth}, \citenamefont {Dooryhee}, \citenamefont
  {Soderholm}, \citenamefont {Cross}, \citenamefont {Mannix},\ and\
  \citenamefont {Ochiai}}]{PhysRevB.71.075115}%
  \BibitemOpen
  \bibfield  {author} {\bibinfo {author} {\bibfnamefont {U.}~\bibnamefont
  {Staub}}, \bibinfo {author} {\bibfnamefont {M.}~\bibnamefont {Shi}}, \bibinfo
  {author} {\bibfnamefont {C.}~\bibnamefont {Schulze-Briese}}, \bibinfo
  {author} {\bibfnamefont {B.~D.}\ \bibnamefont {Patterson}}, \bibinfo {author}
  {\bibfnamefont {F.}~\bibnamefont {Fauth}}, \bibinfo {author} {\bibfnamefont
  {E.}~\bibnamefont {Dooryhee}}, \bibinfo {author} {\bibfnamefont
  {L.}~\bibnamefont {Soderholm}}, \bibinfo {author} {\bibfnamefont {J.~O.}\
  \bibnamefont {Cross}}, \bibinfo {author} {\bibfnamefont {D.}~\bibnamefont
  {Mannix}}, \ and\ \bibinfo {author} {\bibfnamefont {A.}~\bibnamefont
  {Ochiai}},\ }\bibfield  {title} {\enquote {\bibinfo {title} {{Temperature
  dependence of the crystal structure and charge ordering in
  {${\mathrm{Yb}}_{4}{\mathrm{As}}_{3}$}}},}\ }\href {\doibase
  10.1103/PhysRevB.71.075115} {\bibfield  {journal} {\bibinfo  {journal} {Phys.
  Rev. B}\ }\textbf {\bibinfo {volume} {71}},\ \bibinfo {pages} {075115}
  (\bibinfo {year} {2005})}\BibitemShut {NoStop}%
\bibitem [{\citenamefont {Fulde}\ \emph {et~al.}(1995)\citenamefont {Fulde},
  \citenamefont {Schmidt},\ and\ \citenamefont
  {Thalmeier}}]{0295-5075-31-5-6-013}%
  \BibitemOpen
  \bibfield  {author} {\bibinfo {author} {\bibfnamefont {P.}~\bibnamefont
  {Fulde}}, \bibinfo {author} {\bibfnamefont {B.}~\bibnamefont {Schmidt}}, \
  and\ \bibinfo {author} {\bibfnamefont {P.}~\bibnamefont {Thalmeier}},\
  }\bibfield  {title} {\enquote {\bibinfo {title} {{Theoretical Model for the
  Semi-Metal {Yb$_4$As$_3$}}},}\ }\href
  {http://iopscience.iop.org/0295-5075/31/5-6/013} {\bibfield  {journal}
  {\bibinfo  {journal} {EPL (Europhysics Letters)}\ }\textbf {\bibinfo {volume}
  {31}},\ \bibinfo {pages} {323} (\bibinfo {year} {1995})}\BibitemShut
  {NoStop}%
\bibitem [{\citenamefont {Goto}\ and\ \citenamefont
  {L\"uthi}(2003)}]{doi:10.1080/0001873021000057114}%
  \BibitemOpen
  \bibfield  {author} {\bibinfo {author} {\bibfnamefont {T.}~\bibnamefont
  {Goto}}\ and\ \bibinfo {author} {\bibfnamefont {B.}~\bibnamefont {L\"uthi}},\
  }\bibfield  {title} {\enquote {\bibinfo {title} {{Charge ordering, charge
  fluctuations and lattice effects in strongly correlated electron systems}},}\
  }\href {\doibase 10.1080/0001873021000057114} {\bibfield  {journal} {\bibinfo
   {journal} {Advances in Physics}\ }\textbf {\bibinfo {volume} {52}},\
  \bibinfo {pages} {67--118} (\bibinfo {year} {2003})}\BibitemShut {NoStop}%
\bibitem [{\citenamefont {Arguello}\ \emph {et~al.}(2014)\citenamefont
  {Arguello}, \citenamefont {Chockalingam}, \citenamefont {Rosenthal},
  \citenamefont {Zhao}, \citenamefont {Guti\'errez}, \citenamefont {Kang},
  \citenamefont {Chung}, \citenamefont {Fernandes}, \citenamefont {Jia},
  \citenamefont {Millis}, \citenamefont {Cava},\ and\ \citenamefont
  {Pasupathy}}]{PhysRevB.89.235115}%
  \BibitemOpen
  \bibfield  {author} {\bibinfo {author} {\bibfnamefont {C.~J.}\ \bibnamefont
  {Arguello}}, \bibinfo {author} {\bibfnamefont {S.~P.}\ \bibnamefont
  {Chockalingam}}, \bibinfo {author} {\bibfnamefont {E.~P.}\ \bibnamefont
  {Rosenthal}}, \bibinfo {author} {\bibfnamefont {L.}~\bibnamefont {Zhao}},
  \bibinfo {author} {\bibfnamefont {C.}~\bibnamefont {Guti\'errez}}, \bibinfo
  {author} {\bibfnamefont {J.~H.}\ \bibnamefont {Kang}}, \bibinfo {author}
  {\bibfnamefont {W.~C.}\ \bibnamefont {Chung}}, \bibinfo {author}
  {\bibfnamefont {R.~M.}\ \bibnamefont {Fernandes}}, \bibinfo {author}
  {\bibfnamefont {S.}~\bibnamefont {Jia}}, \bibinfo {author} {\bibfnamefont
  {A.~J.}\ \bibnamefont {Millis}}, \bibinfo {author} {\bibfnamefont {R.~J.}\
  \bibnamefont {Cava}}, \ and\ \bibinfo {author} {\bibfnamefont {A.~N.}\
  \bibnamefont {Pasupathy}},\ }\bibfield  {title} {\enquote {\bibinfo {title}
  {{Visualizing the charge density wave transition in
  {$2H$-${\text{NbSe}}_{2}$} in real space}},}\ }\href {\doibase
  10.1103/PhysRevB.89.235115} {\bibfield  {journal} {\bibinfo  {journal} {Phys.
  Rev. B}\ }\textbf {\bibinfo {volume} {89}},\ \bibinfo {pages} {235115}
  (\bibinfo {year} {2014})}\BibitemShut {NoStop}%
\bibitem [{\citenamefont {Ritschel}\ \emph {et~al.}(2015)\citenamefont
  {Ritschel}, \citenamefont {Trinckauf}, \citenamefont {Koepernik},
  \citenamefont {B{\"u}chner}, \citenamefont {Zimmermann}, \citenamefont
  {Berger}, \citenamefont {Joe}, \citenamefont {Abbamonte},\ and\ \citenamefont
  {Geck}}]{ritschel2015orbital}%
  \BibitemOpen
  \bibfield  {author} {\bibinfo {author} {\bibfnamefont {T.}~\bibnamefont
  {Ritschel}}, \bibinfo {author} {\bibfnamefont {J.}~\bibnamefont {Trinckauf}},
  \bibinfo {author} {\bibfnamefont {K.}~\bibnamefont {Koepernik}}, \bibinfo
  {author} {\bibfnamefont {B.}~\bibnamefont {B{\"u}chner}}, \bibinfo {author}
  {\bibfnamefont {M.~v.}\ \bibnamefont {Zimmermann}}, \bibinfo {author}
  {\bibfnamefont {H.}~\bibnamefont {Berger}}, \bibinfo {author} {\bibfnamefont
  {Y.~I.}\ \bibnamefont {Joe}}, \bibinfo {author} {\bibfnamefont
  {P.}~\bibnamefont {Abbamonte}}, \ and\ \bibinfo {author} {\bibfnamefont
  {J.}~\bibnamefont {Geck}},\ }\bibfield  {title} {\enquote {\bibinfo {title}
  {{Orbital textures and charge density waves in transition metal
  dichalcogenides}},}\ }\href@noop {} {\bibfield  {journal} {\bibinfo
  {journal} {Nature physics}\ }\textbf {\bibinfo {volume} {11}},\ \bibinfo
  {pages} {328} (\bibinfo {year} {2015})}\BibitemShut {NoStop}%
\bibitem [{\citenamefont {Ugeda}\ \emph {et~al.}(2016)\citenamefont {Ugeda},
  \citenamefont {Bradley}, \citenamefont {Zhang}, \citenamefont {Onishi},
  \citenamefont {Chen}, \citenamefont {Ruan}, \citenamefont
  {Ojeda-Aristizabal}, \citenamefont {Ryu}, \citenamefont {Edmonds},
  \citenamefont {Tsai} \emph {et~al.}}]{ugeda2016characterization}%
  \BibitemOpen
  \bibfield  {author} {\bibinfo {author} {\bibfnamefont {M.~M.}\ \bibnamefont
  {Ugeda}}, \bibinfo {author} {\bibfnamefont {A.~J.}\ \bibnamefont {Bradley}},
  \bibinfo {author} {\bibfnamefont {Y.}~\bibnamefont {Zhang}}, \bibinfo
  {author} {\bibfnamefont {S.}~\bibnamefont {Onishi}}, \bibinfo {author}
  {\bibfnamefont {Y.}~\bibnamefont {Chen}}, \bibinfo {author} {\bibfnamefont
  {W.}~\bibnamefont {Ruan}}, \bibinfo {author} {\bibfnamefont {C.}~\bibnamefont
  {Ojeda-Aristizabal}}, \bibinfo {author} {\bibfnamefont {H.}~\bibnamefont
  {Ryu}}, \bibinfo {author} {\bibfnamefont {M.~T.}\ \bibnamefont {Edmonds}},
  \bibinfo {author} {\bibfnamefont {H.-Z.}\ \bibnamefont {Tsai}},  \emph
  {et~al.},\ }\bibfield  {title} {\enquote {\bibinfo {title} {{Characterization
  of collective ground states in single-layer {NbSe$_2$}}},}\ }\href@noop {}
  {\bibfield  {journal} {\bibinfo  {journal} {Nature Physics}\ }\textbf
  {\bibinfo {volume} {12}},\ \bibinfo {pages} {92} (\bibinfo {year}
  {2016})}\BibitemShut {NoStop}%
\bibitem [{\citenamefont {Furuno}\ \emph {et~al.}(1988)\citenamefont {Furuno},
  \citenamefont {Ando}, \citenamefont {Kunii}, \citenamefont {Ochiai},
  \citenamefont {Suzuki}, \citenamefont {Fujioka}, \citenamefont {Suzuki},
  \citenamefont {Sasaki},\ and\ \citenamefont {Kasuya}}]{FURUNO1988117}%
  \BibitemOpen
  \bibfield  {author} {\bibinfo {author} {\bibfnamefont {T.}~\bibnamefont
  {Furuno}}, \bibinfo {author} {\bibfnamefont {K.}~\bibnamefont {Ando}},
  \bibinfo {author} {\bibfnamefont {S.}~\bibnamefont {Kunii}}, \bibinfo
  {author} {\bibfnamefont {A.}~\bibnamefont {Ochiai}}, \bibinfo {author}
  {\bibfnamefont {H.}~\bibnamefont {Suzuki}}, \bibinfo {author} {\bibfnamefont
  {M.}~\bibnamefont {Fujioka}}, \bibinfo {author} {\bibfnamefont
  {T.}~\bibnamefont {Suzuki}}, \bibinfo {author} {\bibfnamefont
  {W.}~\bibnamefont {Sasaki}}, \ and\ \bibinfo {author} {\bibfnamefont
  {T.}~\bibnamefont {Kasuya}},\ }\bibfield  {title} {\enquote {\bibinfo {title}
  {{Physical properties of {Sm$_3$Se$_4$} at low temperatures}},}\ }\href
  {\doibase 10.1016/0304-8853(88)90333-2} {\bibfield  {journal} {\bibinfo
  {journal} {Journal of Magnetism and Magnetic Materials}\ }\textbf {\bibinfo
  {volume} {76-77}},\ \bibinfo {pages} {117 -- 118} (\bibinfo {year}
  {1988})}\BibitemShut {NoStop}%
\bibitem [{\citenamefont {Irkhin}\ and\ \citenamefont
  {Katsnelson}(1990)}]{IRKHIN199047}%
  \BibitemOpen
  \bibfield  {author} {\bibinfo {author} {\bibfnamefont {V.~Yu.}\ \bibnamefont
  {Irkhin}}\ and\ \bibinfo {author} {\bibfnamefont {M.~I.}\ \bibnamefont
  {Katsnelson}},\ }\bibfield  {title} {\enquote {\bibinfo {title} {{{RVB}-type
  states in systems with charge and spin degrees of freedom: {Sm$_3$Se$_4$},
  {Y$_{1-x}$Sc$_x$Mn$_2$} etc.}}}\ }\href {\doibase
  10.1016/0375-9601(90)90058-V} {\bibfield  {journal} {\bibinfo  {journal}
  {Physics Letters A}\ }\textbf {\bibinfo {volume} {150}},\ \bibinfo {pages}
  {47 -- 50} (\bibinfo {year} {1990})}\BibitemShut {NoStop}%
\bibitem [{\citenamefont {Wachter}(1980)}]{doi:10.1080/01418638008221893}%
  \BibitemOpen
  \bibfield  {author} {\bibinfo {author} {\bibfnamefont {P.}~\bibnamefont
  {Wachter}},\ }\bibfield  {title} {\enquote {\bibinfo {title} {{Physics of
  {Eu$_3$S$_4$} and {Sm$_3$S$_4$}}},}\ }\href {\doibase
  10.1080/01418638008221893} {\bibfield  {journal} {\bibinfo  {journal}
  {Philosophical Magazine B}\ }\textbf {\bibinfo {volume} {42}},\ \bibinfo
  {pages} {497--498} (\bibinfo {year} {1980})}\BibitemShut {NoStop}%
\bibitem [{\citenamefont {Schlenker}\ and\ \citenamefont
  {Marezio}(1980)}]{doi:10.1080/01418638008221887}%
  \BibitemOpen
  \bibfield  {author} {\bibinfo {author} {\bibfnamefont {C.}~\bibnamefont
  {Schlenker}}\ and\ \bibinfo {author} {\bibfnamefont {M.}~\bibnamefont
  {Marezio}},\ }\bibfield  {title} {\enquote {\bibinfo {title} {{The
  order–disorder transition of {Ti$^{3+}$-Ti$^{3+}$} pairs in {Ti$_4$O$_7$}
  and {(Ti$_{1-x}$V$_x$)$_4$O$_7$}}},}\ }\href {\doibase
  10.1080/01418638008221887} {\bibfield  {journal} {\bibinfo  {journal}
  {Philosophical Magazine B}\ }\textbf {\bibinfo {volume} {42}},\ \bibinfo
  {pages} {453--472} (\bibinfo {year} {1980})}\BibitemShut {NoStop}%
\bibitem [{\citenamefont {Chakraverty}(1980)}]{doi:10.1080/01418638008221888}%
  \BibitemOpen
  \bibfield  {author} {\bibinfo {author} {\bibfnamefont {B.~K.}\ \bibnamefont
  {Chakraverty}},\ }\bibfield  {title} {\enquote {\bibinfo {title} {{Charge
  ordering in {Fe$_3$O$_4$}, {Ti$_4$O$_7$} and bipolarons}},}\ }\href {\doibase
  10.1080/01418638008221888} {\bibfield  {journal} {\bibinfo  {journal}
  {Philosophical Magazine B}\ }\textbf {\bibinfo {volume} {42}},\ \bibinfo
  {pages} {473--478} (\bibinfo {year} {1980})}\BibitemShut {NoStop}%
\bibitem [{\citenamefont {Eyert}\ \emph {et~al.}(2004)\citenamefont {Eyert},
  \citenamefont {Schwingenschlögl},\ and\ \citenamefont
  {Eckern}}]{EYERT2004151}%
  \BibitemOpen
  \bibfield  {author} {\bibinfo {author} {\bibfnamefont {V.}~\bibnamefont
  {Eyert}}, \bibinfo {author} {\bibfnamefont {U.}~\bibnamefont
  {Schwingenschlögl}}, \ and\ \bibinfo {author} {\bibfnamefont
  {U.}~\bibnamefont {Eckern}},\ }\bibfield  {title} {\enquote {\bibinfo {title}
  {{Charge order, orbital order, and electron localization in the Magnéli
  phase {Ti$_4$O$_7$}}},}\ }\href {\doibase 10.1016/j.cplett.2004.04.015}
  {\bibfield  {journal} {\bibinfo  {journal} {Chemical Physics Letters}\
  }\textbf {\bibinfo {volume} {390}},\ \bibinfo {pages} {151 -- 156} (\bibinfo
  {year} {2004})}\BibitemShut {NoStop}%
\bibitem [{\citenamefont {Leonov}\ \emph {et~al.}(2006)\citenamefont {Leonov},
  \citenamefont {Yaresko}, \citenamefont {Antonov}, \citenamefont
  {Schwingenschlögl}, \citenamefont {Eyert},\ and\ \citenamefont
  {Anisimov}}]{0953-8984-18-48-022}%
  \BibitemOpen
  \bibfield  {author} {\bibinfo {author} {\bibfnamefont {I.}~\bibnamefont
  {Leonov}}, \bibinfo {author} {\bibfnamefont {A.~N.}\ \bibnamefont {Yaresko}},
  \bibinfo {author} {\bibfnamefont {V.~N.}\ \bibnamefont {Antonov}}, \bibinfo
  {author} {\bibfnamefont {U.}~\bibnamefont {Schwingenschlögl}}, \bibinfo
  {author} {\bibfnamefont {V.}~\bibnamefont {Eyert}}, \ and\ \bibinfo {author}
  {\bibfnamefont {V.~I.}\ \bibnamefont {Anisimov}},\ }\bibfield  {title}
  {\enquote {\bibinfo {title} {{Charge order and spin-singlet pair formation in
  {Ti$_4$O$_7$}}},}\ }\href {http://stacks.iop.org/0953-8984/18/i=48/a=022}
  {\bibfield  {journal} {\bibinfo  {journal} {Journal of Physics: Condensed
  Matter}\ }\textbf {\bibinfo {volume} {18}},\ \bibinfo {pages} {10955}
  (\bibinfo {year} {2006})}\BibitemShut {NoStop}%
\bibitem [{\citenamefont {Dumas}\ \emph {et~al.}(1980)\citenamefont {Dumas},
  \citenamefont {Schlenker},\ and\ \citenamefont
  {Buder}}]{doi:10.1080/01418638008221890}%
  \BibitemOpen
  \bibfield  {author} {\bibinfo {author} {\bibfnamefont {J.}~\bibnamefont
  {Dumas}}, \bibinfo {author} {\bibfnamefont {C.}~\bibnamefont {Schlenker}}, \
  and\ \bibinfo {author} {\bibfnamefont {R.}~\bibnamefont {Buder}},\ }\bibfield
   {title} {\enquote {\bibinfo {title} {{The vanadium bronzes
  {Na$_x$V$_2$O$_5$-$\beta$}}},}\ }\href {\doibase 10.1080/01418638008221890}
  {\bibfield  {journal} {\bibinfo  {journal} {Philosophical Magazine B}\
  }\textbf {\bibinfo {volume} {42}},\ \bibinfo {pages} {485--486} (\bibinfo
  {year} {1980})}\BibitemShut {NoStop}%
\bibitem [{\citenamefont {Galler}\ \emph {et~al.}(2017)\citenamefont {Galler},
  \citenamefont {Thunstr\"om}, \citenamefont {Gunacker}, \citenamefont
  {Tomczak},\ and\ \citenamefont {Held}}]{PhysRevB.95.115107}%
  \BibitemOpen
  \bibfield  {author} {\bibinfo {author} {\bibfnamefont {A.}~\bibnamefont
  {Galler}}, \bibinfo {author} {\bibfnamefont {P.}~\bibnamefont {Thunstr\"om}},
  \bibinfo {author} {\bibfnamefont {P.}~\bibnamefont {Gunacker}}, \bibinfo
  {author} {\bibfnamefont {J.~M.}\ \bibnamefont {Tomczak}}, \ and\ \bibinfo
  {author} {\bibfnamefont {K.}~\bibnamefont {Held}},\ }\bibfield  {title}
  {\enquote {\bibinfo {title} {{{\it Ab initio} dynamical vertex
  approximation}},}\ }\href {\doibase 10.1103/PhysRevB.95.115107} {\bibfield
  {journal} {\bibinfo  {journal} {Phys. Rev. B}\ }\textbf {\bibinfo {volume}
  {95}},\ \bibinfo {pages} {115107} (\bibinfo {year} {2017})}\BibitemShut
  {NoStop}%
\bibitem [{\citenamefont {{van Loon}}\ \emph {et~al.}(2018)\citenamefont {{van
  Loon}}, \citenamefont {{R{\"o}sner}}, \citenamefont {{Sch{\"o}nhoff}},
  \citenamefont {{Katsnelson}},\ and\ \citenamefont
  {{Wehling}}}]{2017arXiv170705640V}%
  \BibitemOpen
  \bibfield  {author} {\bibinfo {author} {\bibfnamefont {E.~G.~C.~P.}\
  \bibnamefont {{van Loon}}}, \bibinfo {author} {\bibfnamefont
  {M.}~\bibnamefont {{R{\"o}sner}}}, \bibinfo {author} {\bibfnamefont
  {G.}~\bibnamefont {{Sch{\"o}nhoff}}}, \bibinfo {author} {\bibfnamefont
  {M.~I.}\ \bibnamefont {{Katsnelson}}}, \ and\ \bibinfo {author}
  {\bibfnamefont {T.~O.}\ \bibnamefont {{Wehling}}},\ }\bibfield  {title}
  {\enquote {\bibinfo {title} {{Competing Coulomb and electron–phonon
  interactions in {NbS$_2$}}},}\ }\href {\doibase 10.1038/s41535-018-0105-4}
  {\bibfield  {journal} {\bibinfo  {journal} {npj Quantum Materials}\ }\textbf
  {\bibinfo {volume} {3}},\ \bibinfo {pages} {32} (\bibinfo {year}
  {2018})}\BibitemShut {NoStop}%
\bibitem [{\citenamefont {Terletska}\ \emph {et~al.}(2017)\citenamefont
  {Terletska}, \citenamefont {Chen},\ and\ \citenamefont
  {Gull}}]{PhysRevB.95.115149}%
  \BibitemOpen
  \bibfield  {author} {\bibinfo {author} {\bibfnamefont {H.}~\bibnamefont
  {Terletska}}, \bibinfo {author} {\bibfnamefont {T.}~\bibnamefont {Chen}}, \
  and\ \bibinfo {author} {\bibfnamefont {E.}~\bibnamefont {Gull}},\ }\bibfield
  {title} {\enquote {\bibinfo {title} {{Charge ordering and correlation effects
  in the extended Hubbard model}},}\ }\href {\doibase
  10.1103/PhysRevB.95.115149} {\bibfield  {journal} {\bibinfo  {journal} {Phys.
  Rev. B}\ }\textbf {\bibinfo {volume} {95}},\ \bibinfo {pages} {115149}
  (\bibinfo {year} {2017})}\BibitemShut {NoStop}%
\bibitem [{\citenamefont {van Loon}\ \emph
  {et~al.}(2014{\natexlab{b}})\citenamefont {van Loon}, \citenamefont
  {Lichtenstein}, \citenamefont {Katsnelson}, \citenamefont {Parcollet},\ and\
  \citenamefont {Hafermann}}]{PhysRevB.90.235135}%
  \BibitemOpen
  \bibfield  {author} {\bibinfo {author} {\bibfnamefont {E.~G. C.~P.}\
  \bibnamefont {van Loon}}, \bibinfo {author} {\bibfnamefont {A.~I.}\
  \bibnamefont {Lichtenstein}}, \bibinfo {author} {\bibfnamefont {M.~I.}\
  \bibnamefont {Katsnelson}}, \bibinfo {author} {\bibfnamefont
  {O.}~\bibnamefont {Parcollet}}, \ and\ \bibinfo {author} {\bibfnamefont
  {H.}~\bibnamefont {Hafermann}},\ }\bibfield  {title} {\enquote {\bibinfo
  {title} {{Beyond extended dynamical mean-field theory: Dual boson approach to
  the two-dimensional extended Hubbard model}},}\ }\href {\doibase
  10.1103/PhysRevB.90.235135} {\bibfield  {journal} {\bibinfo  {journal} {Phys.
  Rev. B}\ }\textbf {\bibinfo {volume} {90}},\ \bibinfo {pages} {235135}
  (\bibinfo {year} {2014}{\natexlab{b}})}\BibitemShut {NoStop}%
\bibitem [{\citenamefont {Stepanov}\ \emph
  {et~al.}(2016{\natexlab{b}})\citenamefont {Stepanov}, \citenamefont {Huber},
  \citenamefont {van Loon}, \citenamefont {Lichtenstein},\ and\ \citenamefont
  {Katsnelson}}]{PhysRevB.94.205110}%
  \BibitemOpen
  \bibfield  {author} {\bibinfo {author} {\bibfnamefont {E.~A.}\ \bibnamefont
  {Stepanov}}, \bibinfo {author} {\bibfnamefont {A.}~\bibnamefont {Huber}},
  \bibinfo {author} {\bibfnamefont {E.~G. C.~P.}\ \bibnamefont {van Loon}},
  \bibinfo {author} {\bibfnamefont {A.~I.}\ \bibnamefont {Lichtenstein}}, \
  and\ \bibinfo {author} {\bibfnamefont {M.~I.}\ \bibnamefont {Katsnelson}},\
  }\bibfield  {title} {\enquote {\bibinfo {title} {{From local to nonlocal
  correlations: The Dual Boson perspective}},}\ }\href {\doibase
  10.1103/PhysRevB.94.205110} {\bibfield  {journal} {\bibinfo  {journal} {Phys.
  Rev. B}\ }\textbf {\bibinfo {volume} {94}},\ \bibinfo {pages} {205110}
  (\bibinfo {year} {2016}{\natexlab{b}})}\BibitemShut {NoStop}%
\bibitem [{\citenamefont {Ayral}\ \emph {et~al.}(2017)\citenamefont {Ayral},
  \citenamefont {Biermann}, \citenamefont {Werner},\ and\ \citenamefont
  {Boehnke}}]{PhysRevB.95.245130}%
  \BibitemOpen
  \bibfield  {author} {\bibinfo {author} {\bibfnamefont {T.}~\bibnamefont
  {Ayral}}, \bibinfo {author} {\bibfnamefont {S.}~\bibnamefont {Biermann}},
  \bibinfo {author} {\bibfnamefont {P.}~\bibnamefont {Werner}}, \ and\ \bibinfo
  {author} {\bibfnamefont {L.}~\bibnamefont {Boehnke}},\ }\bibfield  {title}
  {\enquote {\bibinfo {title} {{Influence of Fock exchange in combined
  many-body perturbation and dynamical mean field theory}},}\ }\href {\doibase
  10.1103/PhysRevB.95.245130} {\bibfield  {journal} {\bibinfo  {journal} {Phys.
  Rev. B}\ }\textbf {\bibinfo {volume} {95}},\ \bibinfo {pages} {245130}
  (\bibinfo {year} {2017})}\BibitemShut {NoStop}%
\bibitem [{\citenamefont {Ayral}\ and\ \citenamefont
  {Parcollet}(2015)}]{PhysRevB.92.115109}%
  \BibitemOpen
  \bibfield  {author} {\bibinfo {author} {\bibfnamefont {T.}~\bibnamefont
  {Ayral}}\ and\ \bibinfo {author} {\bibfnamefont {O.}~\bibnamefont
  {Parcollet}},\ }\bibfield  {title} {\enquote {\bibinfo {title} {{Mott physics
  and spin fluctuations: A unified framework}},}\ }\href {\doibase
  10.1103/PhysRevB.92.115109} {\bibfield  {journal} {\bibinfo  {journal} {Phys.
  Rev. B}\ }\textbf {\bibinfo {volume} {92}},\ \bibinfo {pages} {115109}
  (\bibinfo {year} {2015})}\BibitemShut {NoStop}%
\bibitem [{\citenamefont {Ayral}\ and\ \citenamefont
  {Parcollet}(2016)}]{PhysRevB.93.235124}%
  \BibitemOpen
  \bibfield  {author} {\bibinfo {author} {\bibfnamefont {T.}~\bibnamefont
  {Ayral}}\ and\ \bibinfo {author} {\bibfnamefont {O.}~\bibnamefont
  {Parcollet}},\ }\bibfield  {title} {\enquote {\bibinfo {title} {{Mott physics
  and spin fluctuations: A functional viewpoint}},}\ }\href {\doibase
  10.1103/PhysRevB.93.235124} {\bibfield  {journal} {\bibinfo  {journal} {Phys.
  Rev. B}\ }\textbf {\bibinfo {volume} {93}},\ \bibinfo {pages} {235124}
  (\bibinfo {year} {2016})}\BibitemShut {NoStop}%
\bibitem [{\citenamefont {Sun}\ and\ \citenamefont
  {Kotliar}(2002)}]{PhysRevB.66.085120}%
  \BibitemOpen
  \bibfield  {author} {\bibinfo {author} {\bibfnamefont {P.}~\bibnamefont
  {Sun}}\ and\ \bibinfo {author} {\bibfnamefont {G.}~\bibnamefont {Kotliar}},\
  }\bibfield  {title} {\enquote {\bibinfo {title} {{Extended dynamical
  mean-field theory and $\mathrm{GW}$ method}},}\ }\href {\doibase
  10.1103/PhysRevB.66.085120} {\bibfield  {journal} {\bibinfo  {journal} {Phys.
  Rev. B}\ }\textbf {\bibinfo {volume} {66}},\ \bibinfo {pages} {085120}
  (\bibinfo {year} {2002})}\BibitemShut {NoStop}%
\bibitem [{\citenamefont {Biermann}\ \emph {et~al.}(2003)\citenamefont
  {Biermann}, \citenamefont {Aryasetiawan},\ and\ \citenamefont
  {Georges}}]{PhysRevLett.90.086402}%
  \BibitemOpen
  \bibfield  {author} {\bibinfo {author} {\bibfnamefont {S.}~\bibnamefont
  {Biermann}}, \bibinfo {author} {\bibfnamefont {F.}~\bibnamefont
  {Aryasetiawan}}, \ and\ \bibinfo {author} {\bibfnamefont {A.}~\bibnamefont
  {Georges}},\ }\bibfield  {title} {\enquote {\bibinfo {title}
  {{First-Principles Approach to the Electronic Structure of Strongly
  Correlated Systems: Combining the {$GW$} Approximation and Dynamical
  Mean-Field Theory}},}\ }\href {\doibase 10.1103/PhysRevLett.90.086402}
  {\bibfield  {journal} {\bibinfo  {journal} {Phys. Rev. Lett.}\ }\textbf
  {\bibinfo {volume} {90}},\ \bibinfo {pages} {086402} (\bibinfo {year}
  {2003})}\BibitemShut {NoStop}%
\bibitem [{\citenamefont {Ruban}\ \emph {et~al.}(2004)\citenamefont {Ruban},
  \citenamefont {Shallcross}, \citenamefont {Simak},\ and\ \citenamefont
  {Skriver}}]{PhysRevB.70.125115}%
  \BibitemOpen
  \bibfield  {author} {\bibinfo {author} {\bibfnamefont {A.~V.}\ \bibnamefont
  {Ruban}}, \bibinfo {author} {\bibfnamefont {S.}~\bibnamefont {Shallcross}},
  \bibinfo {author} {\bibfnamefont {S.~I.}\ \bibnamefont {Simak}}, \ and\
  \bibinfo {author} {\bibfnamefont {H.~L.}\ \bibnamefont {Skriver}},\
  }\bibfield  {title} {\enquote {\bibinfo {title} {{Atomic and magnetic
  configurational energetics by the generalized perturbation method}},}\ }\href
  {\doibase 10.1103/PhysRevB.70.125115} {\bibfield  {journal} {\bibinfo
  {journal} {Phys. Rev. B}\ }\textbf {\bibinfo {volume} {70}},\ \bibinfo
  {pages} {125115} (\bibinfo {year} {2004})}\BibitemShut {NoStop}%
\bibitem [{\citenamefont {Shallcross}\ \emph {et~al.}(2005)\citenamefont
  {Shallcross}, \citenamefont {Kissavos}, \citenamefont {Meded},\ and\
  \citenamefont {Ruban}}]{PhysRevB.72.104437}%
  \BibitemOpen
  \bibfield  {author} {\bibinfo {author} {\bibfnamefont {S.}~\bibnamefont
  {Shallcross}}, \bibinfo {author} {\bibfnamefont {A.~E.}\ \bibnamefont
  {Kissavos}}, \bibinfo {author} {\bibfnamefont {V.}~\bibnamefont {Meded}}, \
  and\ \bibinfo {author} {\bibfnamefont {A.~V.}\ \bibnamefont {Ruban}},\
  }\bibfield  {title} {\enquote {\bibinfo {title} {{An {\it ab initio}
  effective Hamiltonian for magnetism including longitudinal spin
  fluctuations}},}\ }\href {\doibase 10.1103/PhysRevB.72.104437} {\bibfield
  {journal} {\bibinfo  {journal} {Phys. Rev. B}\ }\textbf {\bibinfo {volume}
  {72}},\ \bibinfo {pages} {104437} (\bibinfo {year} {2005})}\BibitemShut
  {NoStop}%
\bibitem [{\citenamefont {Korzhavyi}\ \emph {et~al.}(2009)\citenamefont
  {Korzhavyi}, \citenamefont {Ruban}, \citenamefont {Odqvist}, \citenamefont
  {Nilsson},\ and\ \citenamefont {Johansson}}]{PhysRevB.79.054202}%
  \BibitemOpen
  \bibfield  {author} {\bibinfo {author} {\bibfnamefont {P.~A.}\ \bibnamefont
  {Korzhavyi}}, \bibinfo {author} {\bibfnamefont {A.~V.}\ \bibnamefont
  {Ruban}}, \bibinfo {author} {\bibfnamefont {J.}~\bibnamefont {Odqvist}},
  \bibinfo {author} {\bibfnamefont {J.-O.}\ \bibnamefont {Nilsson}}, \ and\
  \bibinfo {author} {\bibfnamefont {B.}~\bibnamefont {Johansson}},\ }\bibfield
  {title} {\enquote {\bibinfo {title} {{Electronic structure and effective
  chemical and magnetic exchange interactions in bcc {Fe-Cr} alloys}},}\ }\href
  {\doibase 10.1103/PhysRevB.79.054202} {\bibfield  {journal} {\bibinfo
  {journal} {Phys. Rev. B}\ }\textbf {\bibinfo {volume} {79}},\ \bibinfo
  {pages} {054202} (\bibinfo {year} {2009})}\BibitemShut {NoStop}%
\bibitem [{\citenamefont {Ekholm}\ \emph {et~al.}(2010)\citenamefont {Ekholm},
  \citenamefont {Zapolsky}, \citenamefont {Ruban}, \citenamefont {Vernyhora},
  \citenamefont {Ledue},\ and\ \citenamefont
  {Abrikosov}}]{PhysRevLett.105.167208}%
  \BibitemOpen
  \bibfield  {author} {\bibinfo {author} {\bibfnamefont {M.}~\bibnamefont
  {Ekholm}}, \bibinfo {author} {\bibfnamefont {H.}~\bibnamefont {Zapolsky}},
  \bibinfo {author} {\bibfnamefont {A.~V.}\ \bibnamefont {Ruban}}, \bibinfo
  {author} {\bibfnamefont {I.}~\bibnamefont {Vernyhora}}, \bibinfo {author}
  {\bibfnamefont {D.}~\bibnamefont {Ledue}}, \ and\ \bibinfo {author}
  {\bibfnamefont {I.~A.}\ \bibnamefont {Abrikosov}},\ }\bibfield  {title}
  {\enquote {\bibinfo {title} {{Influence of the Magnetic State on the Chemical
  Order-Disorder Transition Temperature in {Fe-Ni} Permalloy}},}\ }\href
  {\doibase 10.1103/PhysRevLett.105.167208} {\bibfield  {journal} {\bibinfo
  {journal} {Phys. Rev. Lett.}\ }\textbf {\bibinfo {volume} {105}},\ \bibinfo
  {pages} {167208} (\bibinfo {year} {2010})}\BibitemShut {NoStop}%
\bibitem [{\citenamefont {Alling}\ \emph {et~al.}(2011)\citenamefont {Alling},
  \citenamefont {Ruban}, \citenamefont {Karimi}, \citenamefont {Hultman},\ and\
  \citenamefont {Abrikosov}}]{PhysRevB.83.104203}%
  \BibitemOpen
  \bibfield  {author} {\bibinfo {author} {\bibfnamefont {B.}~\bibnamefont
  {Alling}}, \bibinfo {author} {\bibfnamefont {A.~V.}\ \bibnamefont {Ruban}},
  \bibinfo {author} {\bibfnamefont {A.}~\bibnamefont {Karimi}}, \bibinfo
  {author} {\bibfnamefont {L.}~\bibnamefont {Hultman}}, \ and\ \bibinfo
  {author} {\bibfnamefont {I.~A.}\ \bibnamefont {Abrikosov}},\ }\bibfield
  {title} {\enquote {\bibinfo {title} {{Unified cluster expansion method
  applied to the configurational thermodynamics of cubic
  {Ti${}_{1\ensuremath{-}x}$Al${}_{x}$N}}},}\ }\href {\doibase
  10.1103/PhysRevB.83.104203} {\bibfield  {journal} {\bibinfo  {journal} {Phys.
  Rev. B}\ }\textbf {\bibinfo {volume} {83}},\ \bibinfo {pages} {104203}
  (\bibinfo {year} {2011})}\BibitemShut {NoStop}%
\bibitem [{\citenamefont {Ruban}\ and\ \citenamefont
  {Abrikosov}(2008)}]{0034-4885-71-4-046501}%
  \BibitemOpen
  \bibfield  {author} {\bibinfo {author} {\bibfnamefont {A.~V.}\ \bibnamefont
  {Ruban}}\ and\ \bibinfo {author} {\bibfnamefont {I.~A.}\ \bibnamefont
  {Abrikosov}},\ }\bibfield  {title} {\enquote {\bibinfo {title}
  {{Configurational thermodynamics of alloys from first principles: effective
  cluster interactions}},}\ }\href
  {http://stacks.iop.org/0034-4885/71/i=4/a=046501} {\bibfield  {journal}
  {\bibinfo  {journal} {Reports on Progress in Physics}\ }\textbf {\bibinfo
  {volume} {71}},\ \bibinfo {pages} {046501} (\bibinfo {year}
  {2008})}\BibitemShut {NoStop}%
\bibitem [{\citenamefont {Connolly}\ and\ \citenamefont
  {Williams}(1983)}]{PhysRevB.27.5169}%
  \BibitemOpen
  \bibfield  {author} {\bibinfo {author} {\bibfnamefont {J.~W.~D.}\
  \bibnamefont {Connolly}}\ and\ \bibinfo {author} {\bibfnamefont {A.~R.}\
  \bibnamefont {Williams}},\ }\bibfield  {title} {\enquote {\bibinfo {title}
  {{Density-functional theory applied to phase transformations in
  transition-metal alloys}},}\ }\href {\doibase 10.1103/PhysRevB.27.5169}
  {\bibfield  {journal} {\bibinfo  {journal} {Phys. Rev. B}\ }\textbf {\bibinfo
  {volume} {27}},\ \bibinfo {pages} {5169--5172} (\bibinfo {year}
  {1983})}\BibitemShut {NoStop}%
\bibitem [{\citenamefont {Hennion}(1983)}]{0305-4608-13-11-017}%
  \BibitemOpen
  \bibfield  {author} {\bibinfo {author} {\bibfnamefont {M.}~\bibnamefont
  {Hennion}},\ }\bibfield  {title} {\enquote {\bibinfo {title} {{Chemical SRO
  effects in ferromagnetic Fe alloys in relation to electronic band
  structure}},}\ }\href {http://stacks.iop.org/0305-4608/13/i=11/a=017}
  {\bibfield  {journal} {\bibinfo  {journal} {Journal of Physics F: Metal
  Physics}\ }\textbf {\bibinfo {volume} {13}},\ \bibinfo {pages} {2351}
  (\bibinfo {year} {1983})}\BibitemShut {NoStop}%
\bibitem [{\citenamefont {Ducastelle}\ and\ \citenamefont
  {Ducastelle}(1991)}]{ducastelle1991order}%
  \BibitemOpen
  \bibfield  {author} {\bibinfo {author} {\bibfnamefont {F.}~\bibnamefont
  {Ducastelle}}\ and\ \bibinfo {author} {\bibfnamefont {F.}~\bibnamefont
  {Ducastelle}},\ }\href@noop {} {\emph {\bibinfo {title} {{Order and phase
  stability in alloys}}}}\ (\bibinfo  {publisher} {North-Holland Amsterdam},\
  \bibinfo {year} {1991})\BibitemShut {NoStop}%
\bibitem [{\citenamefont {Liechtenstein}\ \emph {et~al.}(1984)\citenamefont
  {Liechtenstein}, \citenamefont {Katsnelson},\ and\ \citenamefont
  {Gubanov}}]{LKG84}%
  \BibitemOpen
  \bibfield  {author} {\bibinfo {author} {\bibfnamefont {A.~I.}\ \bibnamefont
  {Liechtenstein}}, \bibinfo {author} {\bibfnamefont {M.~I.}\ \bibnamefont
  {Katsnelson}}, \ and\ \bibinfo {author} {\bibfnamefont {V.~A.}\ \bibnamefont
  {Gubanov}},\ }\bibfield  {title} {\enquote {\bibinfo {title} {{Exchange
  interactions and spin-wave stiffness in ferromagnetic metals}},}\ }\href
  {http://stacks.iop.org/0305-4608/14/i=7/a=007} {\bibfield  {journal}
  {\bibinfo  {journal} {Journal of Physics F: Metal Physics}\ }\textbf
  {\bibinfo {volume} {14}},\ \bibinfo {pages} {L125} (\bibinfo {year}
  {1984})}\BibitemShut {NoStop}%
\bibitem [{\citenamefont {Liechtenstein}\ \emph {et~al.}(1985)\citenamefont
  {Liechtenstein}, \citenamefont {Katsnelson},\ and\ \citenamefont
  {Gubanov}}]{LKG85}%
  \BibitemOpen
  \bibfield  {author} {\bibinfo {author} {\bibfnamefont {A.~I.}\ \bibnamefont
  {Liechtenstein}}, \bibinfo {author} {\bibfnamefont {M.~I.}\ \bibnamefont
  {Katsnelson}}, \ and\ \bibinfo {author} {\bibfnamefont {V.~A.}\ \bibnamefont
  {Gubanov}},\ }\bibfield  {title} {\enquote {\bibinfo {title} {{Local spin
  excitations and Curie temperature of iron}},}\ }\href {\doibase
  http://dx.doi.org/10.1016/0038-1098(85)90007-9} {\bibfield  {journal}
  {\bibinfo  {journal} {Solid State Communications}\ }\textbf {\bibinfo
  {volume} {54}},\ \bibinfo {pages} {327 -- 329} (\bibinfo {year}
  {1985})}\BibitemShut {NoStop}%
\bibitem [{\citenamefont {Liechtenstein}\ \emph {et~al.}(1987)\citenamefont
  {Liechtenstein}, \citenamefont {Katsnelson}, \citenamefont {Antropov},\ and\
  \citenamefont {Gubanov}}]{LKAG87}%
  \BibitemOpen
  \bibfield  {author} {\bibinfo {author} {\bibfnamefont {A.~I.}\ \bibnamefont
  {Liechtenstein}}, \bibinfo {author} {\bibfnamefont {M.~I.}\ \bibnamefont
  {Katsnelson}}, \bibinfo {author} {\bibfnamefont {V.~P.}\ \bibnamefont
  {Antropov}}, \ and\ \bibinfo {author} {\bibfnamefont {V.~A.}\ \bibnamefont
  {Gubanov}},\ }\bibfield  {title} {\enquote {\bibinfo {title} {{Local spin
  density functional approach to the theory of exchange interactions in
  ferromagnetic metals and alloys}},}\ }\href {\doibase
  http://dx.doi.org/10.1016/0304-8853(87)90721-9} {\bibfield  {journal}
  {\bibinfo  {journal} {Journal of Magnetism and Magnetic Materials}\ }\textbf
  {\bibinfo {volume} {67}},\ \bibinfo {pages} {65 -- 74} (\bibinfo {year}
  {1987})}\BibitemShut {NoStop}%
\bibitem [{\citenamefont {Katsnelson}\ and\ \citenamefont
  {Lichtenstein}(2000)}]{KL2000}%
  \BibitemOpen
  \bibfield  {author} {\bibinfo {author} {\bibfnamefont {M.~I.}\ \bibnamefont
  {Katsnelson}}\ and\ \bibinfo {author} {\bibfnamefont {A.~I.}\ \bibnamefont
  {Lichtenstein}},\ }\bibfield  {title} {\enquote {\bibinfo {title}
  {{First-principles calculations of magnetic interactions in correlated
  systems}},}\ }\href {\doibase 10.1103/PhysRevB.61.8906} {\bibfield  {journal}
  {\bibinfo  {journal} {Phys. Rev. B}\ }\textbf {\bibinfo {volume} {61}},\
  \bibinfo {pages} {8906--8912} (\bibinfo {year} {2000})}\BibitemShut {NoStop}%
\bibitem [{\citenamefont {Stepanov}\ \emph {et~al.}(2018)\citenamefont
  {Stepanov}, \citenamefont {Brener}, \citenamefont {Krien}, \citenamefont
  {Harland}, \citenamefont {Lichtenstein},\ and\ \citenamefont
  {Katsnelson}}]{PhysRevLett.121.037204}%
  \BibitemOpen
  \bibfield  {author} {\bibinfo {author} {\bibfnamefont {E.~A.}\ \bibnamefont
  {Stepanov}}, \bibinfo {author} {\bibfnamefont {S.}~\bibnamefont {Brener}},
  \bibinfo {author} {\bibfnamefont {F.}~\bibnamefont {Krien}}, \bibinfo
  {author} {\bibfnamefont {M.}~\bibnamefont {Harland}}, \bibinfo {author}
  {\bibfnamefont {A.~I.}\ \bibnamefont {Lichtenstein}}, \ and\ \bibinfo
  {author} {\bibfnamefont {M.~I.}\ \bibnamefont {Katsnelson}},\ }\bibfield
  {title} {\enquote {\bibinfo {title} {{Effective Heisenberg Model and Exchange
  Interaction for Strongly Correlated Systems}},}\ }\href {\doibase
  10.1103/PhysRevLett.121.037204} {\bibfield  {journal} {\bibinfo  {journal}
  {Phys. Rev. Lett.}\ }\textbf {\bibinfo {volume} {121}},\ \bibinfo {pages}
  {037204} (\bibinfo {year} {2018})}\BibitemShut {NoStop}%
\bibitem [{\citenamefont {Terletska}\ \emph {et~al.}(2018)\citenamefont
  {Terletska}, \citenamefont {Chen}, \citenamefont {Paki},\ and\ \citenamefont
  {Gull}}]{PhysRevB.97.115117}%
  \BibitemOpen
  \bibfield  {author} {\bibinfo {author} {\bibfnamefont {Hanna}\ \bibnamefont
  {Terletska}}, \bibinfo {author} {\bibfnamefont {Tianran}\ \bibnamefont
  {Chen}}, \bibinfo {author} {\bibfnamefont {Joseph}\ \bibnamefont {Paki}}, \
  and\ \bibinfo {author} {\bibfnamefont {Emanuel}\ \bibnamefont {Gull}},\
  }\bibfield  {title} {\enquote {\bibinfo {title} {Charge ordering and nonlocal
  correlations in the doped extended hubbard model},}\ }\href {\doibase
  10.1103/PhysRevB.97.115117} {\bibfield  {journal} {\bibinfo  {journal} {Phys.
  Rev. B}\ }\textbf {\bibinfo {volume} {97}},\ \bibinfo {pages} {115117}
  (\bibinfo {year} {2018})}\BibitemShut {NoStop}%
\bibitem [{\citenamefont {Sengupta}\ and\ \citenamefont
  {Georges}(1995)}]{PhysRevB.52.10295}%
  \BibitemOpen
  \bibfield  {author} {\bibinfo {author} {\bibfnamefont {A.~M.}\ \bibnamefont
  {Sengupta}}\ and\ \bibinfo {author} {\bibfnamefont {A.}~\bibnamefont
  {Georges}},\ }\bibfield  {title} {\enquote {\bibinfo {title}
  {{Non-Fermi-liquid behavior near a $T=0$ spin-glass transition}},}\ }\href
  {\doibase 10.1103/PhysRevB.52.10295} {\bibfield  {journal} {\bibinfo
  {journal} {Phys. Rev. B}\ }\textbf {\bibinfo {volume} {52}},\ \bibinfo
  {pages} {10295--10302} (\bibinfo {year} {1995})}\BibitemShut {NoStop}%
\bibitem [{\citenamefont {Si}\ and\ \citenamefont
  {Smith}(1996)}]{PhysRevLett.77.3391}%
  \BibitemOpen
  \bibfield  {author} {\bibinfo {author} {\bibfnamefont {Q.}~\bibnamefont
  {Si}}\ and\ \bibinfo {author} {\bibfnamefont {J.~L.}\ \bibnamefont {Smith}},\
  }\bibfield  {title} {\enquote {\bibinfo {title} {{Kosterlitz-Thouless
  Transition and Short Range Spatial Correlations in an Extended Hubbard
  Model}},}\ }\href {\doibase 10.1103/PhysRevLett.77.3391} {\bibfield
  {journal} {\bibinfo  {journal} {Phys. Rev. Lett.}\ }\textbf {\bibinfo
  {volume} {77}},\ \bibinfo {pages} {3391--3394} (\bibinfo {year}
  {1996})}\BibitemShut {NoStop}%
\bibitem [{\citenamefont {Smith}\ and\ \citenamefont
  {Si}(2000)}]{PhysRevB.61.5184}%
  \BibitemOpen
  \bibfield  {author} {\bibinfo {author} {\bibfnamefont {J.~L.}\ \bibnamefont
  {Smith}}\ and\ \bibinfo {author} {\bibfnamefont {Q.}~\bibnamefont {Si}},\
  }\bibfield  {title} {\enquote {\bibinfo {title} {{Spatial correlations in
  dynamical mean-field theory}},}\ }\href {\doibase 10.1103/PhysRevB.61.5184}
  {\bibfield  {journal} {\bibinfo  {journal} {Phys. Rev. B}\ }\textbf {\bibinfo
  {volume} {61}},\ \bibinfo {pages} {5184--5193} (\bibinfo {year}
  {2000})}\BibitemShut {NoStop}%
\bibitem [{\citenamefont {Chitra}\ and\ \citenamefont
  {Kotliar}(2000)}]{PhysRevLett.84.3678}%
  \BibitemOpen
  \bibfield  {author} {\bibinfo {author} {\bibfnamefont {R.}~\bibnamefont
  {Chitra}}\ and\ \bibinfo {author} {\bibfnamefont {G.}~\bibnamefont
  {Kotliar}},\ }\bibfield  {title} {\enquote {\bibinfo {title} {{Effect of Long
  Range Coulomb Interactions on the Mott Transition}},}\ }\href {\doibase
  10.1103/PhysRevLett.84.3678} {\bibfield  {journal} {\bibinfo  {journal}
  {Phys. Rev. Lett.}\ }\textbf {\bibinfo {volume} {84}},\ \bibinfo {pages}
  {3678--3681} (\bibinfo {year} {2000})}\BibitemShut {NoStop}%
\bibitem [{\citenamefont {Chitra}\ and\ \citenamefont
  {Kotliar}(2001)}]{PhysRevB.63.115110}%
  \BibitemOpen
  \bibfield  {author} {\bibinfo {author} {\bibfnamefont {R.}~\bibnamefont
  {Chitra}}\ and\ \bibinfo {author} {\bibfnamefont {G.}~\bibnamefont
  {Kotliar}},\ }\bibfield  {title} {\enquote {\bibinfo {title}
  {{Effective-action approach to strongly correlated fermion systems}},}\
  }\href {\doibase 10.1103/PhysRevB.63.115110} {\bibfield  {journal} {\bibinfo
  {journal} {Phys. Rev. B}\ }\textbf {\bibinfo {volume} {63}},\ \bibinfo
  {pages} {115110} (\bibinfo {year} {2001})}\BibitemShut {NoStop}%
\bibitem [{\citenamefont {Metzner}\ and\ \citenamefont
  {Vollhardt}(1989)}]{PhysRevLett.62.324}%
  \BibitemOpen
  \bibfield  {author} {\bibinfo {author} {\bibfnamefont {W.}~\bibnamefont
  {Metzner}}\ and\ \bibinfo {author} {\bibfnamefont {D.}~\bibnamefont
  {Vollhardt}},\ }\bibfield  {title} {\enquote {\bibinfo {title} {{Correlated
  Lattice Fermions in $d=\ensuremath{\infty}$ Dimensions}},}\ }\href {\doibase
  10.1103/PhysRevLett.62.324} {\bibfield  {journal} {\bibinfo  {journal} {Phys.
  Rev. Lett.}\ }\textbf {\bibinfo {volume} {62}},\ \bibinfo {pages} {324--327}
  (\bibinfo {year} {1989})}\BibitemShut {NoStop}%
\bibitem [{\citenamefont {Georges}\ \emph {et~al.}(1996)\citenamefont
  {Georges}, \citenamefont {Kotliar}, \citenamefont {Krauth},\ and\
  \citenamefont {Rozenberg}}]{RevModPhys.68.13}%
  \BibitemOpen
  \bibfield  {author} {\bibinfo {author} {\bibfnamefont {A.}~\bibnamefont
  {Georges}}, \bibinfo {author} {\bibfnamefont {G.}~\bibnamefont {Kotliar}},
  \bibinfo {author} {\bibfnamefont {W.}~\bibnamefont {Krauth}}, \ and\ \bibinfo
  {author} {\bibfnamefont {M.~J.}\ \bibnamefont {Rozenberg}},\ }\bibfield
  {title} {\enquote {\bibinfo {title} {{Dynamical mean-field theory of strongly
  correlated fermion systems and the limit of infinite dimensions}},}\ }\href
  {\doibase 10.1103/RevModPhys.68.13} {\bibfield  {journal} {\bibinfo
  {journal} {Rev. Mod. Phys.}\ }\textbf {\bibinfo {volume} {68}},\ \bibinfo
  {pages} {13--125} (\bibinfo {year} {1996})}\BibitemShut {NoStop}%
\bibitem [{\citenamefont {Hedin}(1965)}]{PhysRev.139.A796}%
  \BibitemOpen
  \bibfield  {author} {\bibinfo {author} {\bibfnamefont {L.}~\bibnamefont
  {Hedin}},\ }\bibfield  {title} {\enquote {\bibinfo {title} {{New Method for
  Calculating the One-Particle Green's Function with Application to the
  Electron-Gas Problem}},}\ }\href {\doibase 10.1103/PhysRev.139.A796}
  {\bibfield  {journal} {\bibinfo  {journal} {Phys. Rev.}\ }\textbf {\bibinfo
  {volume} {139}},\ \bibinfo {pages} {A796--A823} (\bibinfo {year}
  {1965})}\BibitemShut {NoStop}%
\bibitem [{\citenamefont {Aryasetiawan}\ and\ \citenamefont
  {Gunnarsson}(1998)}]{0034-4885-61-3-002}%
  \BibitemOpen
  \bibfield  {author} {\bibinfo {author} {\bibfnamefont {F.}~\bibnamefont
  {Aryasetiawan}}\ and\ \bibinfo {author} {\bibfnamefont {O.}~\bibnamefont
  {Gunnarsson}},\ }\bibfield  {title} {\enquote {\bibinfo {title} {{The {GW}
  method}},}\ }\href {http://stacks.iop.org/0034-4885/61/i=3/a=002} {\bibfield
  {journal} {\bibinfo  {journal} {Reports on Progress in Physics}\ }\textbf
  {\bibinfo {volume} {61}},\ \bibinfo {pages} {237} (\bibinfo {year}
  {1998})}\BibitemShut {NoStop}%
\bibitem [{\citenamefont {Hedin}(1999)}]{0953-8984-11-42-201}%
  \BibitemOpen
  \bibfield  {author} {\bibinfo {author} {\bibfnamefont {L.}~\bibnamefont
  {Hedin}},\ }\bibfield  {title} {\enquote {\bibinfo {title} {{On correlation
  effects in electron spectroscopies and the {GW} approximation}},}\ }\href
  {http://stacks.iop.org/0953-8984/11/i=42/a=201} {\bibfield  {journal}
  {\bibinfo  {journal} {Journal of Physics: Condensed Matter}\ }\textbf
  {\bibinfo {volume} {11}},\ \bibinfo {pages} {R489} (\bibinfo {year}
  {1999})}\BibitemShut {NoStop}%
\bibitem [{\citenamefont {Vonsovsky}\ and\ \citenamefont
  {Katsnelson}(1979)}]{0022-3719-12-11-015}%
  \BibitemOpen
  \bibfield  {author} {\bibinfo {author} {\bibfnamefont {S.~V.}\ \bibnamefont
  {Vonsovsky}}\ and\ \bibinfo {author} {\bibfnamefont {M.~I.}\ \bibnamefont
  {Katsnelson}},\ }\bibfield  {title} {\enquote {\bibinfo {title} {{Some types
  of instabilities in the electron energy spectrum of the polar model of the
  crystal. I. The maximum-polarity state}},}\ }\href
  {http://stacks.iop.org/0022-3719/12/i=11/a=015} {\bibfield  {journal}
  {\bibinfo  {journal} {Journal of Physics C: Solid State Physics}\ }\textbf
  {\bibinfo {volume} {12}},\ \bibinfo {pages} {2043} (\bibinfo {year}
  {1979})}\BibitemShut {NoStop}%
\bibitem [{\citenamefont {Zhuravlev}\ \emph {et~al.}(1997)\citenamefont
  {Zhuravlev}, \citenamefont {Katsnelson},\ and\ \citenamefont
  {Trefilov}}]{PhysRevB.56.12939}%
  \BibitemOpen
  \bibfield  {author} {\bibinfo {author} {\bibfnamefont {A.~K.}\ \bibnamefont
  {Zhuravlev}}, \bibinfo {author} {\bibfnamefont {M.~I.}\ \bibnamefont
  {Katsnelson}}, \ and\ \bibinfo {author} {\bibfnamefont {A.~V.}\ \bibnamefont
  {Trefilov}},\ }\bibfield  {title} {\enquote {\bibinfo {title} {{Electronic
  phase transitions in a one-dimensional spinless fermion model with competing
  interactions}},}\ }\href {\doibase 10.1103/PhysRevB.56.12939} {\bibfield
  {journal} {\bibinfo  {journal} {Phys. Rev. B}\ }\textbf {\bibinfo {volume}
  {56}},\ \bibinfo {pages} {12939--12946} (\bibinfo {year} {1997})}\BibitemShut
  {NoStop}%
\bibitem [{\citenamefont {van Loon}\ \emph {et~al.}(2016)\citenamefont {van
  Loon}, \citenamefont {Krien}, \citenamefont {Hafermann}, \citenamefont
  {Stepanov}, \citenamefont {Lichtenstein},\ and\ \citenamefont
  {Katsnelson}}]{PhysRevB.93.155162}%
  \BibitemOpen
  \bibfield  {author} {\bibinfo {author} {\bibfnamefont {E.~G. C.~P.}\
  \bibnamefont {van Loon}}, \bibinfo {author} {\bibfnamefont {F.}~\bibnamefont
  {Krien}}, \bibinfo {author} {\bibfnamefont {H.}~\bibnamefont {Hafermann}},
  \bibinfo {author} {\bibfnamefont {E.~A.}\ \bibnamefont {Stepanov}}, \bibinfo
  {author} {\bibfnamefont {A.~I.}\ \bibnamefont {Lichtenstein}}, \ and\
  \bibinfo {author} {\bibfnamefont {M.~I.}\ \bibnamefont {Katsnelson}},\
  }\bibfield  {title} {\enquote {\bibinfo {title} {{Double occupancy in
  dynamical mean-field theory and the dual boson approach}},}\ }\href {\doibase
  10.1103/PhysRevB.93.155162} {\bibfield  {journal} {\bibinfo  {journal} {Phys.
  Rev. B}\ }\textbf {\bibinfo {volume} {93}},\ \bibinfo {pages} {155162}
  (\bibinfo {year} {2016})}\BibitemShut {NoStop}%
\bibitem [{\citenamefont {Rohringer}\ \emph {et~al.}(2012)\citenamefont
  {Rohringer}, \citenamefont {Valli},\ and\ \citenamefont
  {Toschi}}]{PhysRevB.86.125114}%
  \BibitemOpen
  \bibfield  {author} {\bibinfo {author} {\bibfnamefont {G.}~\bibnamefont
  {Rohringer}}, \bibinfo {author} {\bibfnamefont {A.}~\bibnamefont {Valli}}, \
  and\ \bibinfo {author} {\bibfnamefont {A.}~\bibnamefont {Toschi}},\
  }\bibfield  {title} {\enquote {\bibinfo {title} {{Local electronic
  correlation at the two-particle level}},}\ }\href {\doibase
  10.1103/PhysRevB.86.125114} {\bibfield  {journal} {\bibinfo  {journal} {Phys.
  Rev. B}\ }\textbf {\bibinfo {volume} {86}},\ \bibinfo {pages} {125114}
  (\bibinfo {year} {2012})}\BibitemShut {NoStop}%
\bibitem [{\citenamefont {{Y. M. Vilk}}\ and\ \citenamefont {{A.-M. S.
  Tremblay}}(1997)}]{refId0}%
  \BibitemOpen
  \bibfield  {author} {\bibinfo {author} {\bibnamefont {{Y. M. Vilk}}}\ and\
  \bibinfo {author} {\bibnamefont {{A.-M. S. Tremblay}}},\ }\bibfield  {title}
  {\enquote {\bibinfo {title} {{Non-Perturbative Many-Body Approach to the
  Hubbard Model and Single-Particle Pseudogap}},}\ }\href {\doibase
  10.1051/jp1:1997135} {\bibfield  {journal} {\bibinfo  {journal} {J. Phys. I
  France}\ }\textbf {\bibinfo {volume} {7}},\ \bibinfo {pages} {1309--1368}
  (\bibinfo {year} {1997})}\BibitemShut {NoStop}%
\bibitem [{\citenamefont {Antropov}\ \emph {et~al.}(1995)\citenamefont
  {Antropov}, \citenamefont {Katsnelson}, \citenamefont {van Schilfgaarde},\
  and\ \citenamefont {Harmon}}]{PhysRevLett.75.729}%
  \BibitemOpen
  \bibfield  {author} {\bibinfo {author} {\bibfnamefont {V.~P.}\ \bibnamefont
  {Antropov}}, \bibinfo {author} {\bibfnamefont {M.~I.}\ \bibnamefont
  {Katsnelson}}, \bibinfo {author} {\bibfnamefont {M.}~\bibnamefont {van
  Schilfgaarde}}, \ and\ \bibinfo {author} {\bibfnamefont {B.~N.}\ \bibnamefont
  {Harmon}},\ }\bibfield  {title} {\enquote {\bibinfo {title} {{{\it Ab Initio}
  Spin Dynamics in Magnets}},}\ }\href {\doibase 10.1103/PhysRevLett.75.729}
  {\bibfield  {journal} {\bibinfo  {journal} {Phys. Rev. Lett.}\ }\textbf
  {\bibinfo {volume} {75}},\ \bibinfo {pages} {729--732} (\bibinfo {year}
  {1995})}\BibitemShut {NoStop}%
\end{thebibliography}%

\end{document}